\documentclass[12pt]{article}
\usepackage{amsmath}
\usepackage{amssymb}
\usepackage{amsthm}
\usepackage{amsfonts}
\usepackage[colorlinks=true,citecolor=blue,urlcolor=blue,pdfstartview=FitB]{hyperref}
\usepackage[top=1.1in, bottom=1.1in, left=1in, right=1in]{geometry}
\usepackage[doublespacing]{setspace}
\usepackage{natbib}
\usepackage{algorithm}
\usepackage{algpseudocode}
\let\Algorithm\algorithm
\renewcommand\algorithm[1][]{\Algorithm[#1]\setstretch{1.15}}

\newtheorem{theorem}{Theorem}[]

\newtheorem{corollary}{Corollary}[]

\DeclareMathOperator*{\argmin}{\arg\!\min}
\DeclareMathOperator*{\argmax}{\arg\!\max}

\newcommand{\ms}{\boldsymbol}
\newcommand{\mb}{\mathbf}

\usepackage{color}
\usepackage{graphicx}
\usepackage{caption}
\usepackage{subcaption}

\begin{document}
\title{\vspace{-1.5cm} Post-Lasso Inference for High-Dimensional Regression}
\author{X. Jessie Jeng\footnote{Address for correspondence: X. Jessie Jeng,  Department of Statistics, North Carolina State University, Raleigh, NC, USA. Email: \texttt{xjjeng@ncsu.edu}.}  
\ Huimin Peng\footnote{Department of Statistics, North Carolina State University, Raleigh, NC, USA} 
\ and Wenbin Lu\footnote{Department of Statistics, North Carolina State University, Raleigh, NC, USA} 
}
\date{}
\maketitle
\begin{abstract}
Among the most popular variable selection procedures in high-dimensional regression, Lasso provides a solution path to rank the variables and determines a cut-off position on the path to select variables and estimate coefficients. In this paper, we consider variable selection from a new perspective motivated by the frequently occurred phenomenon that relevant variables are not completely distinguishable from noise variables on the solution path. We propose to characterize the positions of the first noise variable and the last relevant variable on the path. We then develop a new variable selection procedure to control over-selection of the noise variables ranking after the last relevant variable, and, at the same time, retain a high proportion of relevant variables ranking before the first noise variable. Our procedure utilizes the recently developed covariance test statistic and Q statistic in post-selection inference. In numerical examples, our method compares favorably with other existing methods in selection accuracy and the ability to interpret its results. 

\end{abstract}

\section{Introduction} \label{Introduction}

We consider the linear regression model 
\begin{eqnarray}
\mb{y}=\mb{X}\ms{\beta}^*+\ms{\epsilon}, \qquad \ms{\epsilon}\sim N(0,\sigma^2 \mb{I}),
\label{modellasso}
\end{eqnarray}
where $\mb{y}$ is a vector of response with length $n$, $\mb{X}$ is the $n\times p$ design matrix of standardized predictors, and $\ms{\beta}^*$ a sparse vector of coefficients. In high-dimensional regression, $p$ can be greater than $n$.  Among the most popular methods for variable selection and estimation in the high-dimensional regression, Lasso \citep{ff3} solves the following optimization problem 
\begin{eqnarray} \label{def:lasso}
\hat{\ms{\beta}}(\lambda) =\argmin_{\beta\in\mathcal{R}^{p}} \frac{1}{2} \Vert \mb{y}-\mb{X}\ms{\beta}\Vert_2^2+
\lambda \Vert\ms{\beta}\Vert_1,
\end{eqnarray}
where $\lambda\geq 0$ is a tuning parameter. Lasso provides a solution path, which is the plot of the estimate $\hat{\beta}(\lambda)$ versus the tuning parameter $\lambda$. 
Lasso solution path is piecewise linear with each knot corresponding to the entry of a variable into the selected set. 
Knots are denoted by
$\lambda_1\geq \lambda_2\geq \cdots\geq \lambda_m\geq 0$, 
where $m=\min(n-1,p)$ is the length of the solution path \citep{b5}. 

Recent developments in high-dimensional regression focus on  hypothesis testing for variable selection. Impressive progress has been made in \cite{b20}, \cite{b3}, \cite{b2}, \cite{Barber15}, \cite{Bogdan15}, \cite{Lee16}, \cite{Gsell16}, etc. Specifically, innovative test statistics based on Lasso solution path have been proposed. For example, \cite{b2} construct the covariance test statistic as follows. Along the solution path, the variable indexed by $j_k$ enters the selected model at knot $\lambda_k$.
Define active set right before knot $\lambda_k$ as 
$A_k=\{j_1,j_2,\cdots,j_{k-1}\}$.
In addition, define true active set to be
$A^*=\{j:~\beta^*_j\neq 0\}$ and the size of true active set as $s=\vert A^*\vert$. 
\cite{b2} considers to test the null hypothesis $H_{0k}: ~A^*\subset A_{k}$ conditional upon the active set $A_k$ at knot $\lambda_k$ and propose the covariance test statistic as
\begin{eqnarray}
T_k=\left(\langle \mb{y}, \mb{X}\hat{\ms{\beta}}(\lambda_{k+1})\rangle-
\langle \mb{y}, \mb{X}_{A_k}\tilde{\ms{\beta}}_{A_k}(\lambda_{k+1})\rangle\right)/\sigma^2,
\label{def:covstat}
\end{eqnarray}
where 
$\hat{\ms{\beta}}(\lambda_{k+1})=\argmin_{\beta\in\mathcal{R}^{p}}
\frac{1}{2}\Vert \mb{y}-\mb{X} \ms{\beta} \Vert_2^2+\lambda_{k+1}\Vert\ms{\beta}\Vert_1$ and $ \tilde{\ms{\beta}}_{A_k}(\lambda_{k+1})=\argmin_{\beta\in\mathcal{R}^{\vert A_k\vert}}
\frac{1}{2}\Vert \mb{y}-\mb{X}_{A_k}\ms{\beta}_{A_k}\Vert_2^2+\lambda_{k+1}\Vert\ms{\beta}_{A_k}\Vert_1$.
\citet{b2} derived that under orthogonal design, if all $s$ relevant variables rank ahead of noise variables with probability tending to 1, then for any fixed $d$,
\begin{eqnarray}
(T_{s+1},T_{s+2},\cdots,T_{s+d})
\overset{d}{\to} (\text{Exp}(1),\text{Exp}(1/2),\cdots,\text{Exp}(1/d)),
\label{covdis}
\end{eqnarray}
as $p\to\infty$, and that $T_{1},T_{2},\cdots,T_{d}$ are asymptotically independent . 

Later, \citet{Gsell16} proposed to perform sequential test on 
$H_{0k}: A^*\subset A_{k}$ for $k$ increasing from $0$ to $m$ 
and developed the Q statistics for a stopping rule. The Q statistics are defined as 
\begin{eqnarray} \label{def:Q}
q_k=\text{exp} \left(-\sum_{j=k}^m T_j \right)
\label{qstat}
\end{eqnarray}
for $k = 1, \ldots, m$. It has been proved that if all $s$ relevant variables rank ahead of noise variables and 
$T_{s+1},\cdots,T_{m}$ are independently distributed as $(T_{s+1},\cdots,T_{m}) \sim  (\text{Exp}(1),\text{Exp}(1/2), \\
\cdots,\text{Exp}(1/(m-s)))$,  then
\begin{equation} \label{q_dist}
q_k \overset{d}{=}  k\text{th order statistic of~} m-s \text{~independent standard uniform random variables}
\end{equation}
for $s+1 \le k \le m$.
\citet{Gsell16} developed a stopping rule (TailStop) implementing the Q statistics in the procedure of \cite{wsc19}.  Given the distribution of $q_k$ in (\ref{q_dist}), TailStop controls false discovery rate at a pre-specified level when all relevant variables rank before noise variables on the solution path.

In this paper, we consider more general scenarios where relevant variables and noise variables are not perfectly separated and some  (or all) relevant variables intertwine with noise variables 
on the Lasso solution path.  
Such scenarios would occur when the effect sizes of relevant variables are not large enough. In fact, even with infinitely large effect size, perfect separation on solution path is still not guaranteed  when the number of relevant variables is relatively large \citep{Wainwright:2009, Su17}. 
Studies in theory and method are limited in such general scenarios because the inseparability among relevant and noise variables make it difficult to estimate Type I and/or Type II errors.  
In order to perform variable selection in the more general and realistic settings, we propose a new theoretical framework to characterize the region on the solution path where relevant and noise variables are not distinguishable.      

\begin{figure}[!htbp]
	\centering
	\caption{An example of $m_0$ and $m_1$ on Lasso solution path. 
		$m_0$ is the entry right before first noise variable. 
		$m_1$ is the entry of the last relevant variable.}
	\includegraphics[width=0.75\textwidth,height=0.17\textheight]{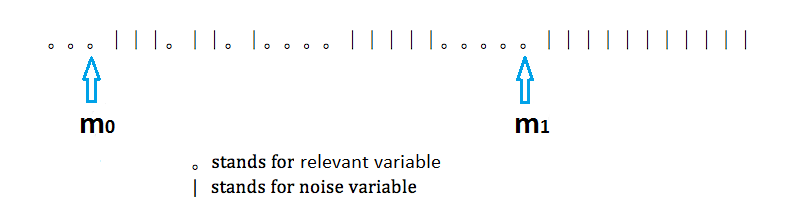}
	\label{fig:path}
\end{figure}

Figure \ref{fig:path} illustrates the indistinguishable region on solution path. $m_0$ represents the position right before the first noise variable on the path such that all entries up to $m_0$ correspond to relevant variables; and $m_1$ represents the position of the last relevant variable where all entries afterwards correspond to noise variables.  Given a solution path, both $m_0$ and $m_1$ are realized but unknown, and the region between $m_0$ and $m_1$ is referred to as the indistinguishable region. 

\vspace{-0.25in}

Given the solution path, a sensible variable selection procedure would select all variables up to $m_0$ but no variables after $m_1$. Since both $m_0$ and $m_1$ are unknown stochastic quantities, the selection procedure should automatically adapt to the unknown $m_0$ and $m_1$.

We develop a new variable selection procedure utilizing the Q statistic, which we refer to as Q-statistic Variable Selection (QVS). 
The new procedure searches through the solution path and determines a cut-off position that is asymptotically between $m_0$ and $m_1$.
In other words,  with probability tending to 1, a high proportion of the relevant variables ranked up to $m_0$ are selected and no noise variables ranked after $m_1$ are included.  
QVS is adaptive to the unknown  indistinguishable region. In the ideal case with perfect separation, i.e. $m_0 = m_1$, QVS consistently estimate the separation point and select all and only the relevant variables. 


\section{Method and Theory}
\label{sec:method_theory}

QVC is inspired by earlier works on estimating the proportion of non-null component in a mixture model of $p$-value \citep{e1,h1}. We extend the technique to high-dimensional regression considering the general scenarios with indistinguishable relevant and noise variables.  

Given a Lasso solution path, QVS searches through the empirical process of the Q statistics: $k/m -q_k-c_m\sqrt{q_k(1-q_k)}, 1\le k\le m$; and determines the cut-off position as  
\begin{eqnarray}
\hat{k}_{qvs}=m \cdot \max_{1\leq k\leq m/2} \left\{
\frac{k}{m}-q_k-c_m\sqrt{q_k(1-q_k)}\right\},
\label{QVS_def}
\end{eqnarray}
where $q_k$ is defined in (\ref{qstat}) and 
$c_m$ is a bounding sequence to control over selection of noise variables after $m_1$.
$c_m$ is constructed as follows. 
For $0<t<1$, let
\[
U_m(t)=\frac{1}{m}\sum_{i=1}^m 1(U_i\leq t),
\]
where $U_1,\cdots,U_m$ are i.i.d.  standard uniform random variables.
Further, let 
\begin{equation} \label{Vm}
V_m = \sup_{t \in (0,1)} \frac{U_m(t)-t}{\sqrt{t(1-t)}}.
\end{equation}
Then determine $c_m$ so that $P(V_m > c_m) <\alpha_m \to 0$  as $m \to \infty$. We set $\alpha_m=1/\sqrt{\log m}$ in this paper.

We consider the general setting where some relevant variables intertwine with noise variables on the Lasso solution path. Denote the positions of noise variables on the solution path as $a_1, \ldots, a_{m-s}$. 
Define 
\[
T^0_{a_k} = \left(\langle \mb{y}, \mb{X}\hat{\ms{\beta}}(\lambda_{a_{k+1}})\rangle-
\langle \mb{y}, \mb{X}_{A_{a_k}}\tilde{\ms{\beta}}_{A_{a_k}}(\lambda_{a_{k+1}})\rangle\right)/\sigma^2,
\] 
where 
$\hat{\ms{\beta}}(\lambda_{a_{k+1}})=\argmin_{\beta\in\mathcal{R}^{p}}
\frac{1}{2}\Vert \mb{y}-\mb{X} \ms{\beta} \Vert_2^2+\lambda_{a_{k+1}}\Vert\ms{\beta}\Vert_1$ and $ \tilde{\ms{\beta}}_{A_{a_k}}(\lambda_{a_{k+1}})=\argmin_{\beta\in\mathcal{R}^{\vert A_{a_k}\vert}}
\frac{1}{2}\Vert \mb{y}-\mb{X}_{A_{a_k}}\ms{\beta}_{A_{a_k}}\Vert_2^2+\lambda_{a_{k+1}}\Vert\ms{\beta}_{A_{a_k}}\Vert_1$. Under orthogonal design, $T^0_{a_k} = \lambda_{a_k}(\lambda_{a_k} - \lambda_{a_{k+1}})$. Note that $T_{a_k}^0$ not necessarily equals to $T_{a_k}$ unless the variable at position $a_k+1$ is a noise variable so that $a_k+ 1 = a_{k+1}$. Since $\lambda_{a_k+1} \ge \lambda_{a_k}+1$, we have $T^0_{a_k} \ge T_{a_k}$. Further, it is easy to see that $T_{a_k}^0 = T_{a_k}$ for $a_k > m_1$. 
We assume that under orthogonal design, 
$T_{a_1}^0,\cdots,T_{a_{m-s}}^0$ are independently distributed as 
\begin{equation} \label{cond:T}
(T_{a_1}^0,\cdots,T_{a_{m-s}}^0) \sim  (\text{Exp}(1),\text{Exp}(1/2),\cdots,\text{Exp}(1/(m-s))). 
\end{equation} 
Recall that \citet{Gsell16} presents a similar condition on  $T_{s+1},\cdots,T_{m}$ assuming all $s$ relevant variables rank ahead of noise variables. Such condition on $T_k$ is not feasible in our case with indistinguishable variables. Our condition (\ref{cond:T}) is more general and includes the condition in \citet{Gsell16} as a special case when $m_1 = s$. 

QVS is constructed upon Q statistics in (\ref{def:Q}). However, unlike in \citet{Gsell16}, $q_k$ is not necessarily distributed as the $k$-th order statistic of $m-s$ independent standard uniform random variables for $s+1 \le k \le m$. In the more general setting, our first theoretical result shows that $\hat{k}_{qvs}$ in (\ref{QVS_def}) provides a lower bound for the unknown $m_1$ in Figure \ref{fig:path}. 

\begin{theorem} \label{thm:lowerbd}
	Consider the stopping rule in (\ref{QVS_def}) under condition (\ref{cond:T}) and $s = o(m)$. Define $m_1$ as the position of the last relevant variable on Lasso solution path (Figure \ref{fig:path}). Then, as $m\to\infty$,  
	\[
	P(\hat{k}_{qvs} \leq m_1) \to 1.
	\]
\end{theorem}
The proof of Theorem \ref{thm:lowerbd} is presented in Section \ref{sec:proof_1}.  Theorem \ref{thm:lowerbd} implies that QVS  
provides a parsimonious variable selection such that  noise variables ranked after $m_1$ are not likely to be over-selected. 

Besides the control of over-selection of noise variables, QVS guarantees to select a high proportion of relevant variables ranked before $m_0$. Our next result shows that QVS provides an asymptotic upper bound for the unknown $m_0$. 

\begin{theorem} \label{thm:upperbd_m0}
	Consider the stopping rule in (\ref{QVS_def}) under condition (\ref{cond:T}) and $s = o(m)$. Define $m_0$ as the position right before the first noise variable on Lasso solution path (Figure \ref{fig:path}). 
	Assume $m_0 \gg \sqrt{\log m}$ with probability tending to 1. Then, as $m\to\infty$,  
	\begin{equation} \label{5}
	P(\hat{k}_{qvs} \geq(1-\epsilon) m_0) \to 1
	\end{equation}
	for any small constant $\epsilon>0$. 
\end{theorem}

The proof of Theorem \ref{thm:upperbd_m0} is provided in Section \ref{sec:proof_2}. 
The upper bound result in Theorem \ref{thm:upperbd_m0} implies that QVS can asymptotically retain a high proportion of relevant variables ranked up to $m_0$. 

Next, we show that under some additional condition on the number of noise variables ranked before $m_1$, which equals to $m_1-s$, $\hat{k}_{qvs}$ consistently estimates the position of the last relevant variable on the solution path. 

\begin{theorem} \label{thm:upperbd_m1}
	Assume the conditions in Theorem \ref{thm:lowerbd}. Additionally, assume $m_1 \gg \sqrt{\log m}$ and $(m_1-s)/m_1 \ll 1$ with probability tending to 1. Then
	\[
	\hat{k}_{qvs}/m_1 \to 1 
	\]
	in probability.
\end{theorem}

The condition $(m_1-s)/m_1 = o_p(1)$ says that the proportion of noise variables ranked before $m_1$ is asymptotically negligible. With this additional condition, QVS can consistently estimate $m_1$ to retain a high proportion of all relevant variables. 

Our last theoretical result is for the special case when $m_0=m_1$ on solution path. 
\cite{Gsell16} has exclusively considered this case in theory on the control of false discovery rate by the TailStop procedure. In our theoretical context, it is straightforward to show that QVS consistently estimate the position of $m_0 (= m_1)$ and select all and only the $s$ relevant variables in this special case. 

\begin{corollary} \label{cor:perfect}
	Consider the stopping rule in (\ref{QVS_def}) under condition (\ref{cond:T}). If $\sqrt{\log m} \ll s \ll m$ and $m_0=m_1=s$ with probability tending to 1. Then
	\[
	\hat{k}_{qvs}/s \to 1 
	\]
	in probability.
\end{corollary}

\section{Simulation}
\label{Simulation}

In our simulation study, design matrix $\mb{X}_{n\times p}$ is a Gaussian random matrix with each row generated from $N_p(0,\ms{\Sigma})$.
Response $\mb{y}$ is generated from $N_n(\mb{X} \ms{\beta}^*, \mb{I})$, where
$\ms{\beta}^*$ is the vector of true coefficients. The locations of non-zero entries of $\ms{\beta}^*$ are randomly simulated. 

For the QVS procedure, we simulate the bounding sequence $c_m$ by the following steps. 
We generate $\mb{X}_{n\times p}$ and $\mb{y}_{n\times 1}$ under the null model and compute the Lasso solution path using the {\it lars} package in r.  Covariance test statistics and Q statistics $\{q_i\}_{i=1}^m$ are calculated by (\ref{def:covstat}) and (\ref{def:Q}), respectively. 
Then, we compute $V_m$ using $V_m=\max_{1\leq i\leq m/2} (i/m-q_i)/\sqrt{q_i(1-q_i)}$.
We repeat the above steps for $1000$ times and obtain $V_m^1,V_m^2,\cdots,V_m^{1000}$. 
The bounding sequence $c_m$ is computed as the upper $\alpha_m$ percentile of $V_m^1,V_m^2,\cdots,V_m^{1000}$ with $\alpha_m=1/\sqrt{\log m}$. 
For each combination of sample size $n$ and dimension $p$, we only need to simulate the bounding sequence once. 

\subsection{Relationships among $\hat{k}_{qvs}$, $m_0$, and $m_1$}

Recall the definitions of the $m_0$ and $m_1$ on the Lasso solution path and Figure \ref{fig:path}.  Table \ref{tab:frequency} reports the realized values of $\hat{k}_{qvs}$, $m_0$, $m_1$, and their relationships. It can be seen that the distance between $m_0$ and $m_1$ increases as the number of relevant variables $s$ increases and/or the dimension $p$ increases. 
$\hat{k}_{qvs}$ is greater than $m_0$ in all cases meaning that all relevant variables ranked up to $m_0$ are retained. On the other hand,  $\hat{k}_{qvs}$ is less than $m_1$ with high frequency meaning that  $\hat{k}_{qvs}$ mostly avoids over-selecting the noise variables after $m_1$. These numerical results support our theoretical findings in Theorem \ref{thm:lowerbd} and \ref{thm:upperbd_m0}. 

\begin{table}[!htbp]
	\centering
	\caption{Mean values of the QVS cut-off position ($\hat{k}_{qvs}$), the position right before the first noise variable on Lasso solution path ($m_0$), and the position of the last relevant variable ($m_1$). Standard errors are in parenthesis. 	
		$F(\hat{k}_{qvs} \ge m_0)$ and $F(\hat{k}_{qvs}\leq m_1)$ represent the frequencies of $\hat{k}_{qvs} \ge m_0$ and $\hat{k}_{qvs}\leq m_1$ in $100$ replications, respectively. 
		In these examples, $n=200$, $Cov({X})={I}$, and non-zero $\beta^*=0.3$.}
	\label{tab:frequency}
	\begin{tabular}{|l|l|l|l|l|l|l|}
		\hline
		p & s & $\hat{k}_{qvs}$ & $m_0$ & $m_1$ & $F(\hat{k}_{qvs}\ge m_0)$ & $F(\hat{k}_{qvs}\leq m_1)$   \\ \hline
		2000 & 10 & 19.60(5.20)  & 3.47(1.92) & 65.53(46.66)  & 1.00 & 0.91  \\   
		& 20 & 39.15(6.14) & 3.24(2.24) & 131.36(40.95) & 1.00 & 1.00 \\  
		& 30 & 54.44(6.58) & 2.71(2.13) & 160.47(29.62) & 1.00 & 1.00 \\  
		& 40 & 66.98(6.65) & 2.45(2.08) & 173.52(22.42) & 1.00 & 1.00  \\ \hline
		10000 & 10 & 27.12(6.55) & 2.00(1.49) & 96.59(50.82) & 1.00 & 0.93  \\ 
		& 20 & 50.67(7.20) & 1.63(1.43) & 141.18(41.50) & 1.00 & 0.98  \\  
		& 30 & 67.48(7.27) & 1.24(1.25) & 157.21(33.52) & 1.00 & 0.99 \\  
		& 40 & 78.11(6.48) & 0.94(1.09) & 163.74(29.26) & 1.00 & 0.99  \\ \hline
	\end{tabular}
\end{table}

\subsection{Comparisons with other methods} \label{sec:simulation_compare}

We compare the performance of QVS with other variable selection methods, such as Lasso with BIC (``BIC"), Lasso with $10$-fold cross-validation (``LCV"), 
Bonferroni procedure applied to the Q statistics (``Q-BON"), and 
Benjamini-Hochberg procedure applied to the Q statistics (``Q-FDR"). Specifically, Q-BON and  Q-FDR select the top-ranked variables on the solution path with sizes equal to $\argmax_k \left\{k: ~ q_k\leq 0.05/m\right\}$ and $\argmax_k \left\{k: ~ q_k\leq 0.05k/m\right\}$, respectively. The nominal levels for both Q-BON and Q-FDR are set at $0.05$. 
We note that Q-FDR is equivalent to the TailStop method introduced in \cite{Gsell16}. 

We demonstrate the performances of these methods by presenting the true positive proportion (TPP), false discovery proportion (FDP), and g-measure of these methods. TPP is the ratio of true positives to the number of relevant variables entered the solution path. FDP is the ratio of false positives to the number of selected variables. TPP and FDP measure the power and type I error of a selection method, respectively. We also compute the g-measure, which is the geometric mean of specificity and sensitivity, i.e.  
g-measure $=\sqrt{\text{specificity}\times \text{sensitivity}}$, where $\text{specificity}$ is the ratio of true negatives to the number of noise variables in the solution path and sensitivity is equivalent to TPP. G-measure can be used to evaluate the overall performance of a variable selection method. Higher value of g-measure indicates better performance \citep{Powers11}. 

\begin{figure}[htpb]
	\includegraphics[width=0.32\textwidth,height=0.32\textheight]{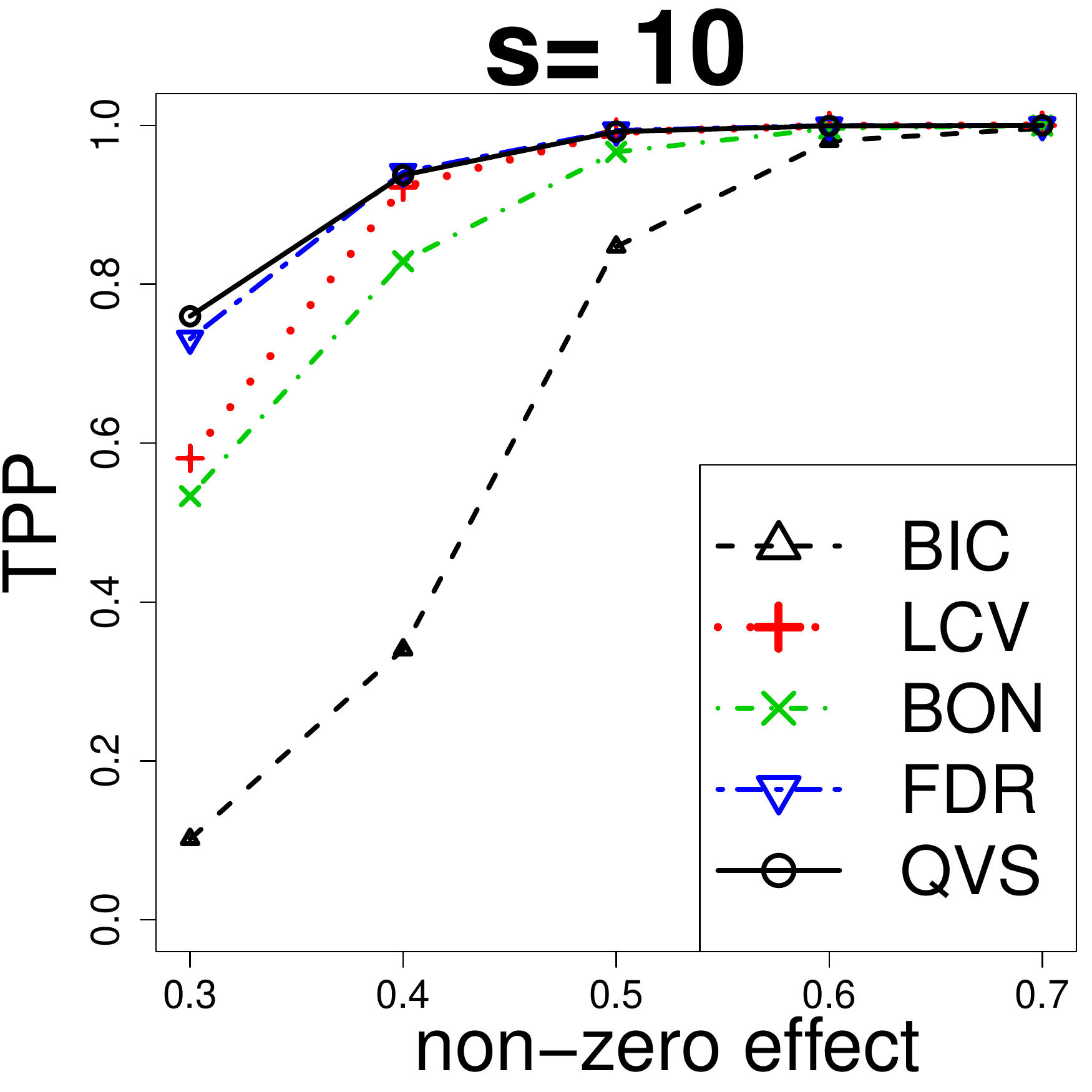}
	\includegraphics[width=0.32\textwidth,height=0.32\textheight]{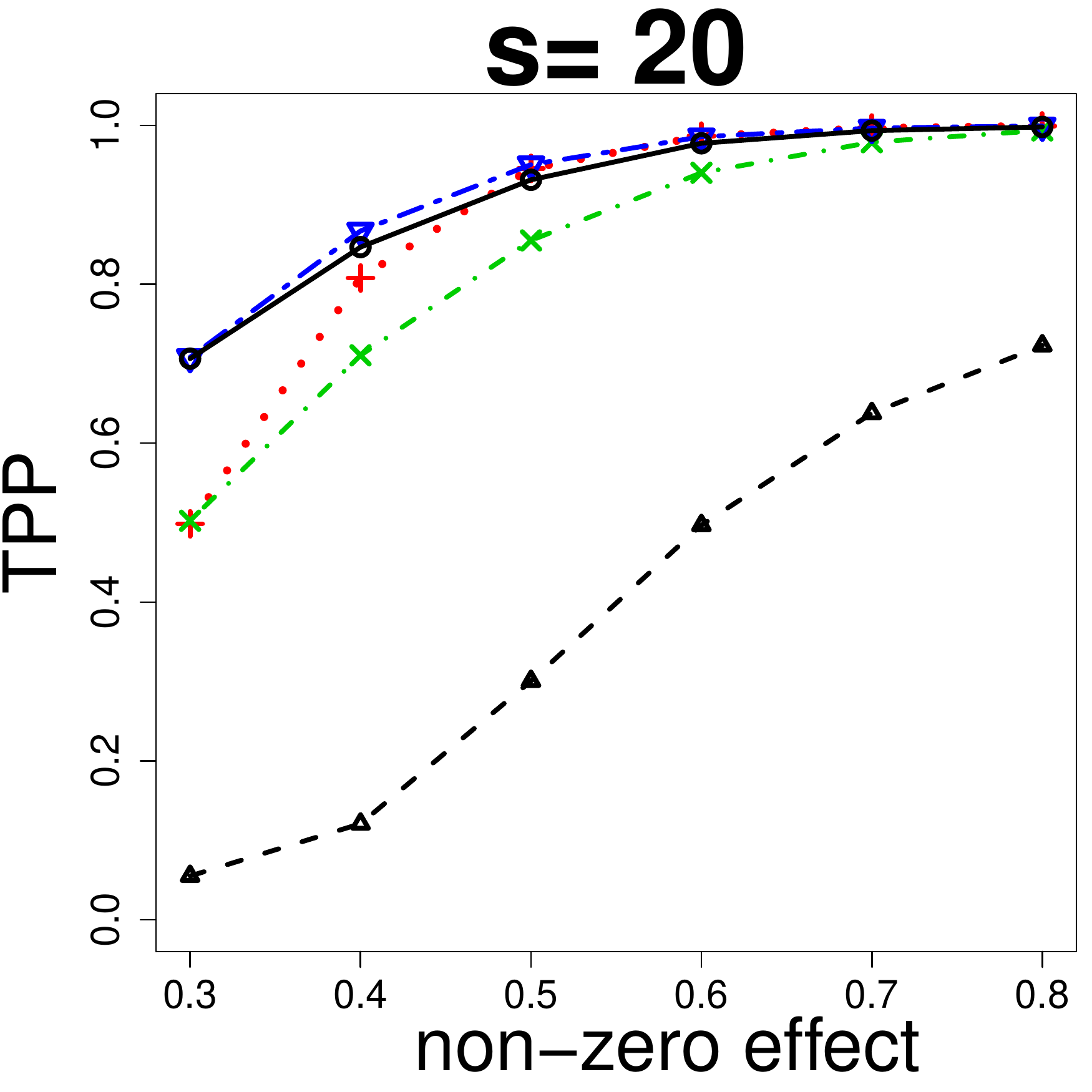}
	\includegraphics[width=0.32\textwidth,height=0.32\textheight]{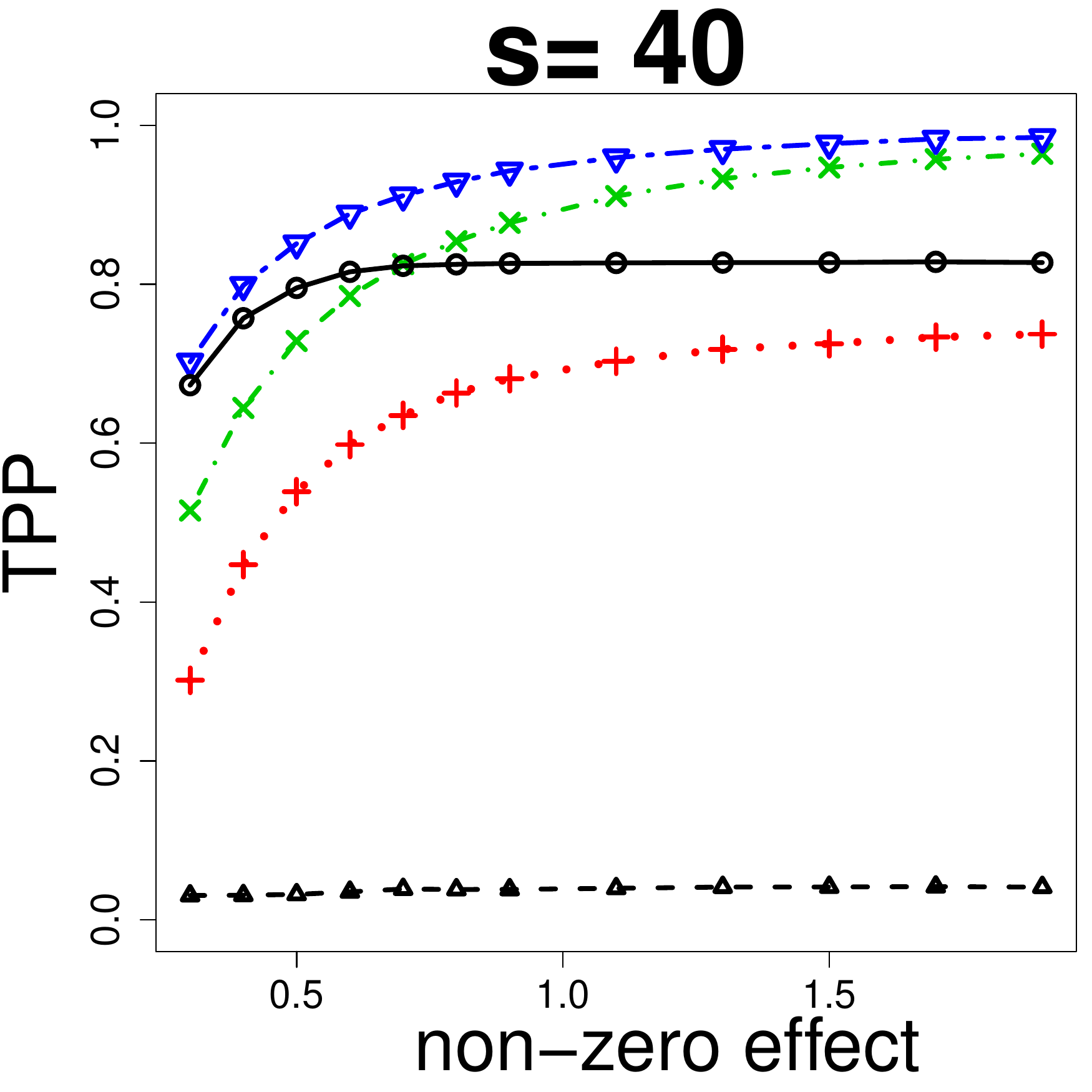}
	\includegraphics[width=0.32\textwidth,height=0.32\textheight]{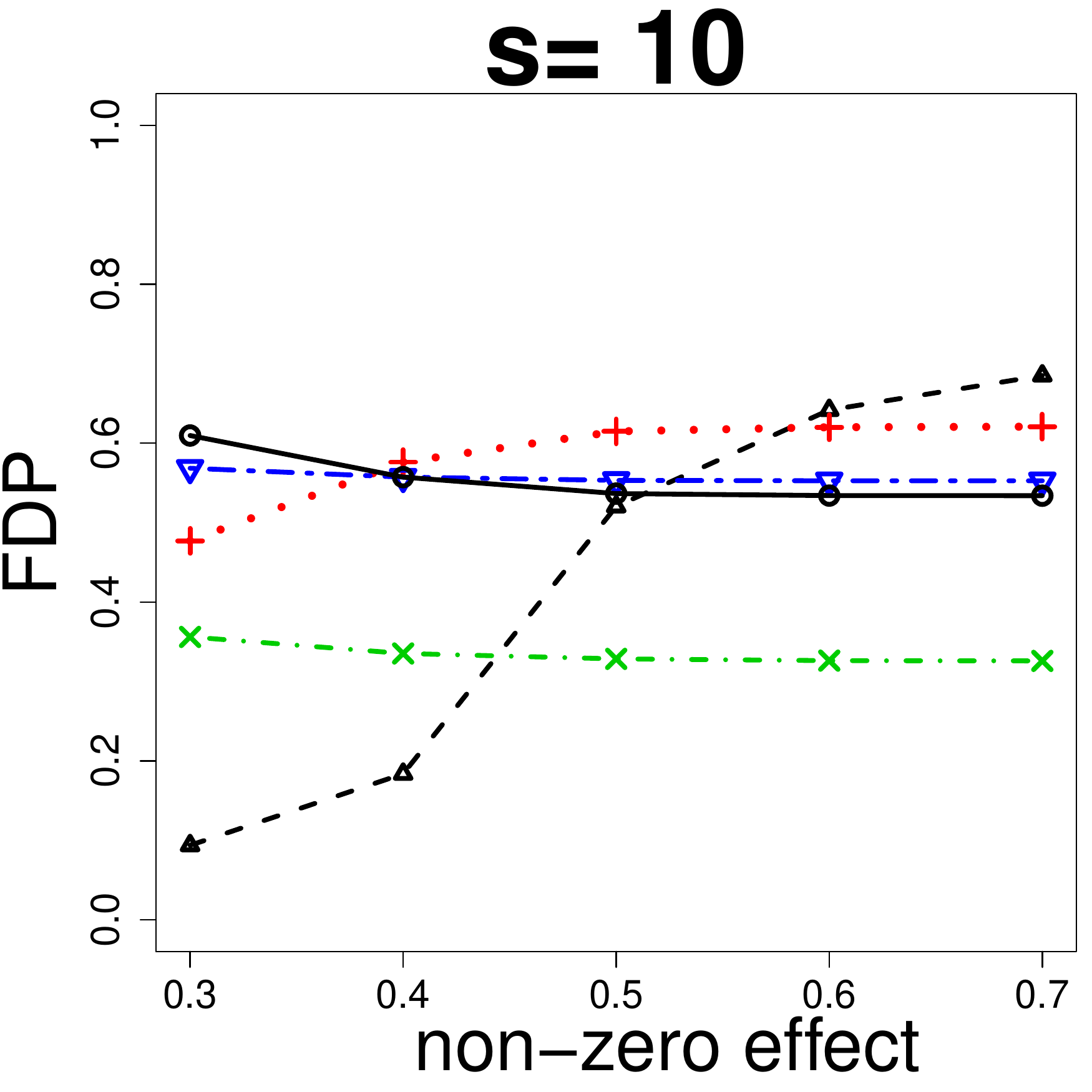}
	\includegraphics[width=0.32\textwidth,height=0.32\textheight]{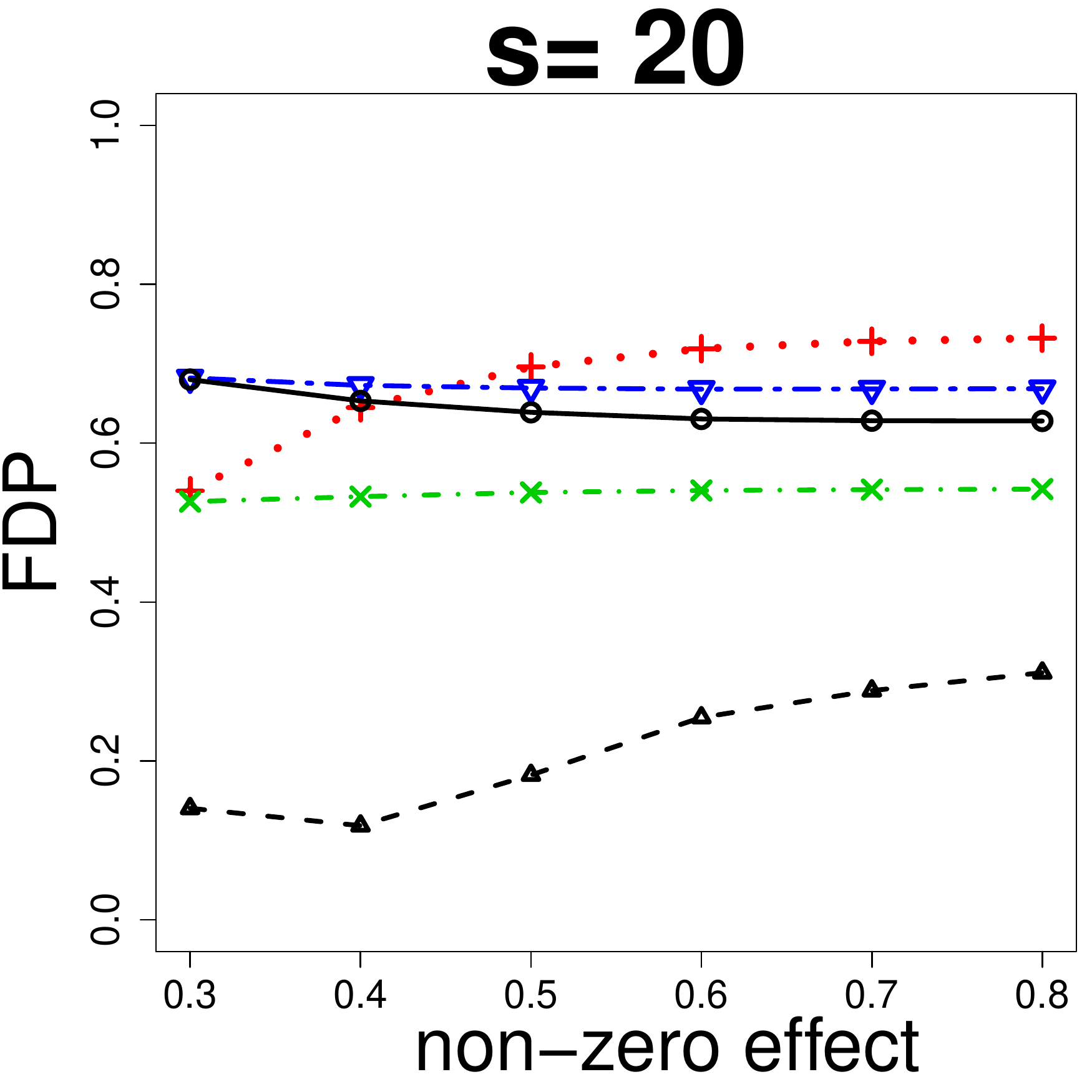}
	\includegraphics[width=0.32\textwidth,height=0.32\textheight]{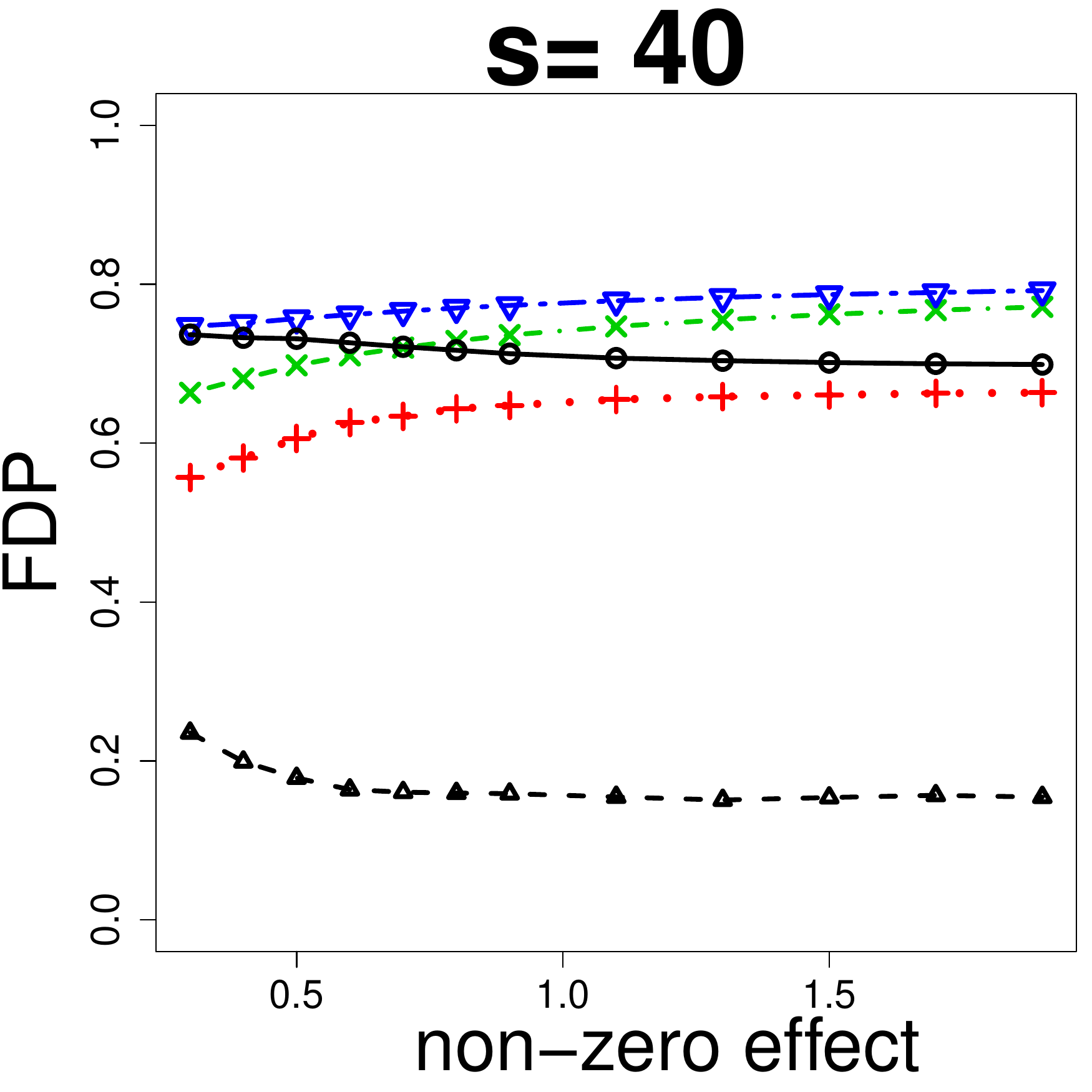}
	\includegraphics[width=0.32\textwidth,height=0.32\textheight]{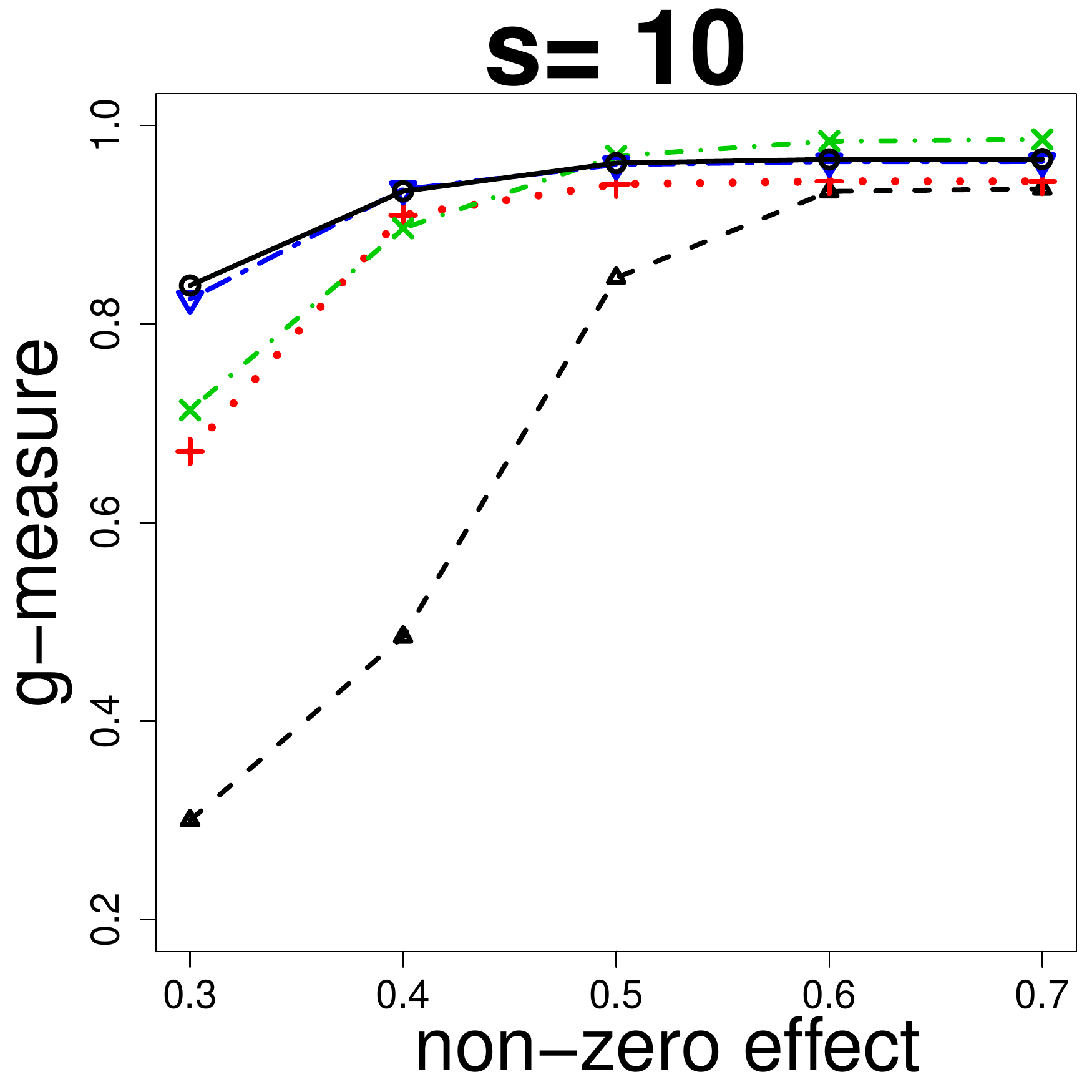}
	\includegraphics[width=0.32\textwidth,height=0.32\textheight]{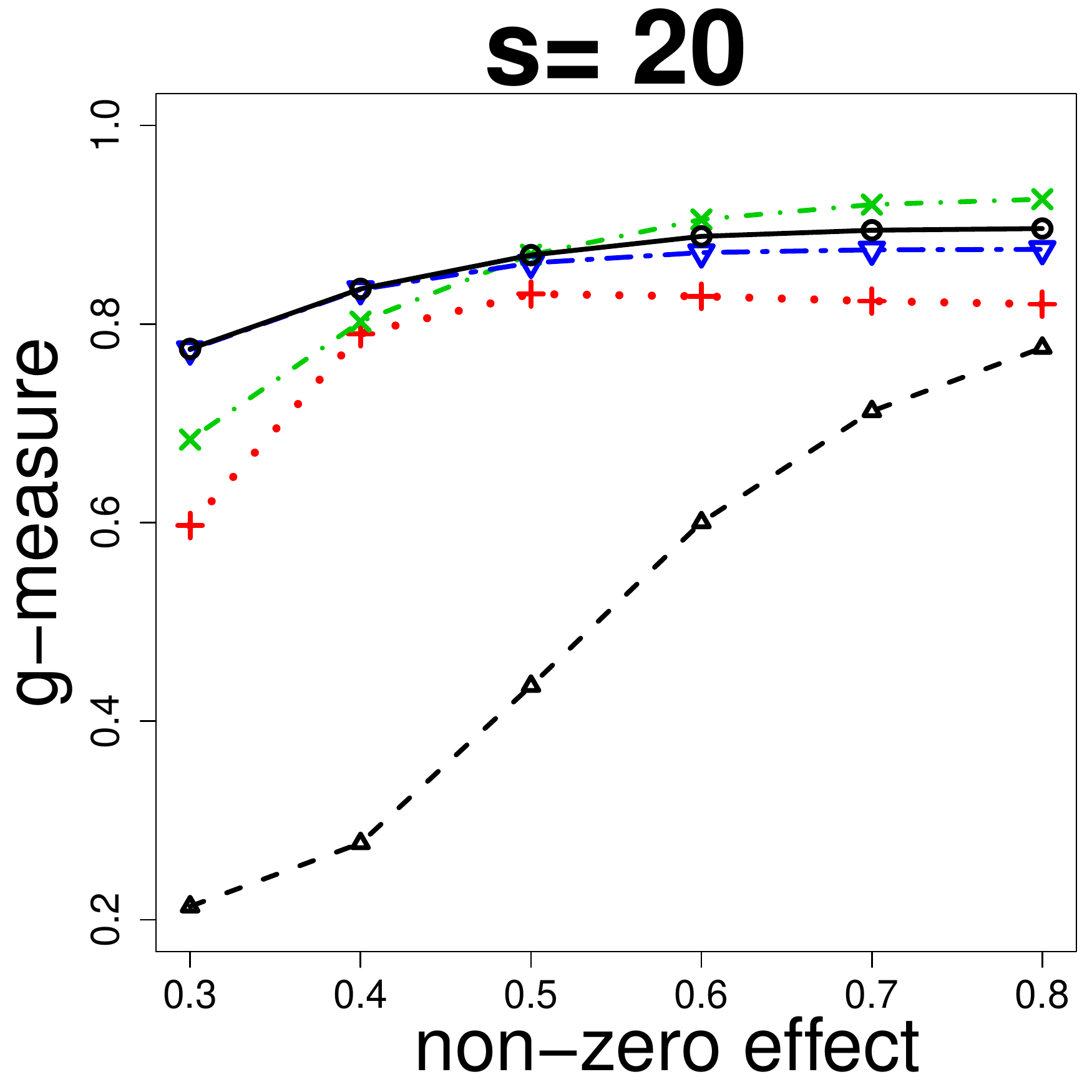}
	\includegraphics[width=0.32\textwidth,height=0.32\textheight]{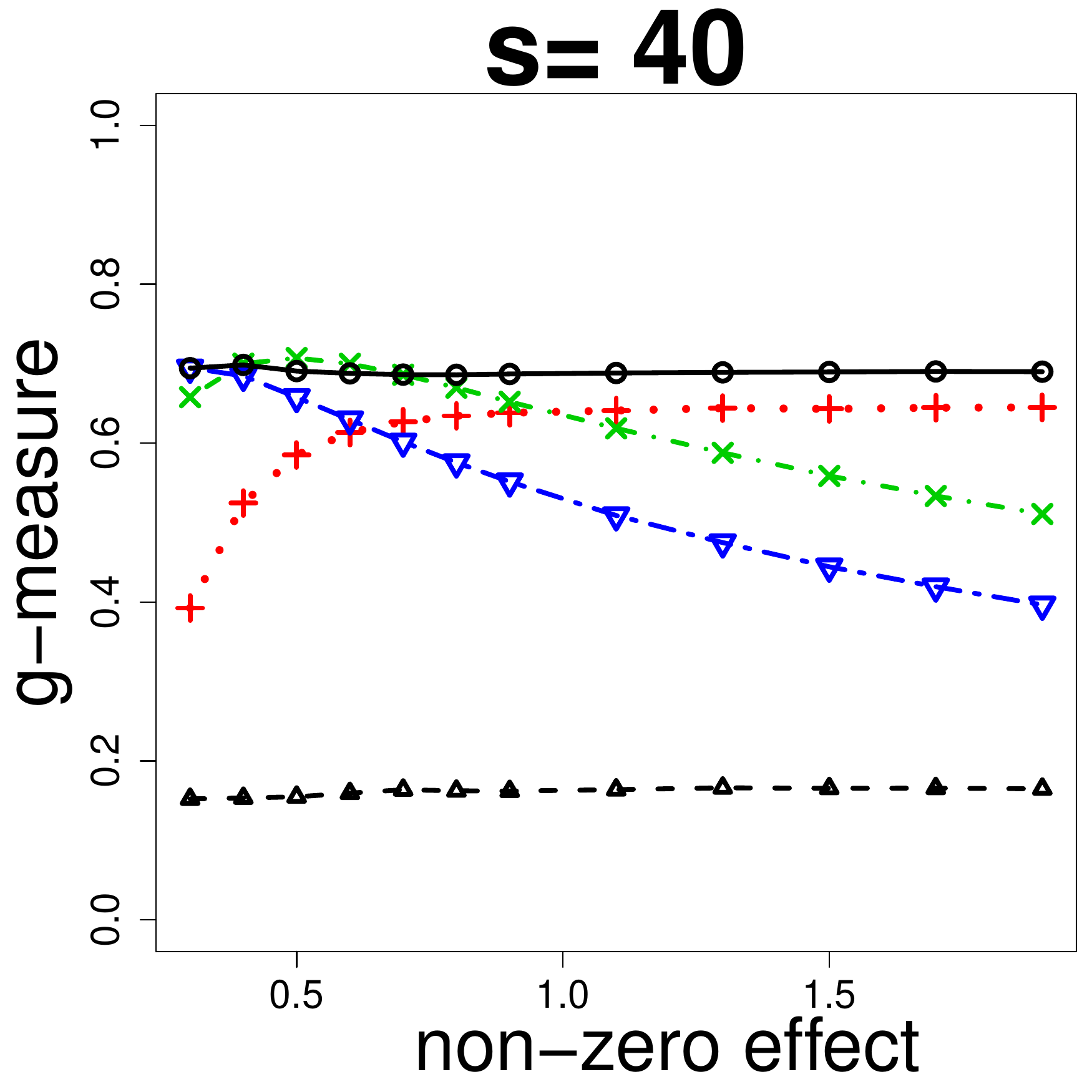}
	\caption{Comparisons of QVS with other methods when $p=2000$, $\ms{\Sigma}=(0.5^{|i-j|})_{i=1,\cdots,p;~j=1,\cdots,p}$, and $n=200$.}
	\label{p2000}
\end{figure}

Figure \ref{p2000} summarizes the mean values of TPP, FDP, and g-measure for different methods under various model settings with $p=2000$, $n=200$ and $Cov(\mb{X}) = \ms{\Sigma}=(0.5^{|i-j|})_{i=1,\cdots,p;~j=1,\cdots,p}$. 
The number of non-zero coefficients $s = 10, 20, 40$, and the non-zero effect vary from 0.3 to 2. 
It can be seen that the Lasso-based BIC and LCV tend to select fewer variables, which results in lower TPP and FDP. On the other hand, the Q Statistic-based methods, Q-BON, Q-FDR, and QVS, all have higher TPP and FDP. However, in these examples, Q-BON does not control family-wise error at the nominal level of $0.05$, and Q-FDR does not control FDR at the nominal of $0.05$.  
The reason is because relevant and noise variables are not perfectly separated in these examples. As illustrated in Table \ref{tab:frequency}, $m_0$ is much smaller than $m_1$, and the results of Q-BON and Q-FDR  cannot be interpreted presumably. In terms of g-measure, QVS generally outperforms other methods.

More simulation results with $p = 400$ and $10000$ are presented in Section \ref{sec:add_sim}.

\section{Real Application}
\label{Application}

We obtain a dataset for expression quantitative trait loci (eQTL) analysis related to Down Syndrome. 
Down Syndrome is one of the most common gene-associated diseases. 
Our dataset includes the expression levels of gene CCT8, which contains a critical region of Down syndrome, and genome-wide single-nucleotide polymorphism (SNP) data from three different populations \citep{g1}: Asia (Japan and China) with sample size $n=90$, Yoruba with $n=60$, and Europe with $n=60$. We perform eQTL mapping to identify SNPs that are potentially associated with the expression level of gene CCT8 for each population. Due to the limited sample size, we randomly select subsets of SNPs with $p=6000, 2000, 4000$ for the three populations, respectively.   

For the sample of each population, we first compute the covariance test statistics by (\ref{def:covstat}) and Q statistics by (\ref{def:Q}) based on Lasso solution path. Table \ref{tab:Qstat_data} presents these statistics for the top 10 knots on the path. 

\begin{table}[!htbp]
	\centering
	\caption{Covariance test statistics and Q statistics along Lasso solution path for samples from Asian, Yoruba, and European populations, respectively. }
	\label{tab:Qstat_data}
	\begin{tabular}{|l|l|l|l|l|l|l|}
		\hline
		knots &  \multicolumn{2}{l|}{Asian} & \multicolumn{2}{l|}{Yoruba} & \multicolumn{2}{l|}{European} \\ \cline{2-7} 
		& Covtest  & Q statistic & Covtest  & Q statistic & Covtest  & Q statistic \\ \hline
		1 & 6.09e-01 & 0.00 & 4.20e-01 &  0.00  & 9.81e-03 & 0.85 \\
		2 & 3.05e 00 &  0.01 & 6.46e 00 &  0.00 & 2.94e-02& 0.85\\ 
		3 & 1.98e-01 &  0.26 & 4.41e-02 &  0.45 & 1.64e-02  & 0.88 \\	
		4 & 5.66e-02 &0.32 & 1.23e-01 &  0.47 & 1.78e-03 & 0.89 \\
		5 & 1.24e-01 &  0.34 & 1.06e-04 & 0.54 & 1.05e-02  & 0.90 \\
		6 & 3.94e-03 & 0.39 & 2.77e-02 &  0.54 & 6.99e-03 & 0.91 \\ 
		7 & 2.05e-02 &  0.39 & 1.60e-03 & 0.55 & 1.06e-03  & 0.91 \\ 
		8 & 2.24e-02 & 0.40 & 4.31e-02 &  0.55 & 6.49e-03 & 0.91 \\ 
		9 & 1.09e-02 & 0.41 & 9.26e-03 &  0.58 & 1.03e-04 & 0.92 \\ 
		10& 4.61e-02 & 0.41 & 1.29e-02 &  0.58 & 7.83e-03  & 0.92 \\ \hline
	\end{tabular}%
\end{table}

We apply QVS as well as all the other methods analyzed in Section \ref{sec:simulation_compare} to the datasets. Table \ref{tab:data_selection} presents the number of selected variables along the solution path for different methods. It can be seen that QVS generally selects more variables than the other methods for these datasets. Particularly, when there exists a gap in the list of Q-statistics, such as for Asian and Yoruba samples, QVS tends to stop right after the gap. This is because such gap is likely to occur in the indistinguishable region between $m_0$ and $m_1$. Stopping right after the gap would include relevant variables ranked before $m_0$ and, at the same time, not over-select the noise variables ranked after $m_1$.       

\begin{table}[ht]
	\centering
	\caption{The number of selected variables long the Lasso solution path for different methods.}
	\label{tab:data_selection}
	\begin{tabular}{|l|l|l|l|l|l|l|l|}
		\hline
		Population & n & p & BIC & LCV & Q-BON & Q-FDR & QVS \\ \hline
		Asian & 90 & 6000  & 0 & 1 & 0 & 0 & 3 \\ 
		Yoruba & 60 & 2000 & 1 & 2 & 2 & 2 & 3 \\ 
		European &60 & 4000 & 0 & 0 & 0 & 0 & 0 \\ \hline
	\end{tabular}
\end{table}

We further verify the result of QVS by comparing with the findings in literature.  \citet{g1} studied the same samples for eQTL mapping but only focused on cis-eQTLs. Therefore, the numbers of SNPs included in their analysis are much smaller with $p= 1955, 1978, 2146$ for the three populations, respectively. More SNP variables are identified in \citet{g1} for each population due to larger ratio of sample size to dimension. 
Table \ref{tab:data_reference} reports the locations on the solution path of the variables identified in \citet{g1}. 
Note that the Lasso solution path is computed using our datasets with lower ratio of sample size to dimension. 
We utilize this result as a reference to evaluation the results of QVS.  

For the solution path of Asian population, according to \citet{g1}, the first noise variable enters after two relevant variables and the last relevant variable enters at the position 46. Therefore, $m_0=2$ and $m_1 = 46$. QVS selects the top 3 variables on the path, which is in-between $m_0$ and $m_1$. This result supports the theoretical property of QVS as a sensible variable selection procedure. Similar results are observed in the other two cases.

\begin{table}[ht]
	\centering
	\caption{Locations on the Lasso solution paths of the reference variables identified in \citet{g1}.} \label{tab:data_reference}
	\begin{tabular}{|l|l|l|l|l|l|l|}
		\hline
		Population & n & p & location of reference variables & $m_0$ & $m_1$ & QVS \\ \hline
		Asian  & 90 & 1955 & 1,~2,~6,~46 & 2 & 46 & 3 \\ 
		Yoruba & 60 & 1978 & 1,~4,~17,~34,~50,~58 & 1 & 58 & 3 \\ 
		European & 60 & 2146 & 2,~4,~18,~30 & 0 & 30 & 0\\ \hline
	\end{tabular}
\end{table}

\section{Conclusion and Discussion}

We consider variable selection in high-dimensional regression from a new perspective motivated by the natural phenomenon that relevant and noise variables are often not completely distinguishable due to high-dimensionality and limited sample size. We demonstrate such phenomenon on Lasso solution path and propose to characterize the positions of the first noise variable and the last relevant variable on the path. We then develop a new variable selection procedure whose result is interpretable in the general scenarios where indistinguishable region exists on the path. The theoretical findings in the paper are very different from the existing results which consider variable selection properties in the ideal setting where all relevant variables rank ahead of noise variables on the solution path. Our new analytic framework is unconventional but highly relevant to Big Data applications. 

The proposed QVS procedure utilizes the Q statistic developed in \cite{Gsell16} that is built upon the limiting distribution of the covariance test statistic developed in \cite{b2}. 
The theoretical analysis in the paper has focused on orthogonal design. In a more general setting where design matrix is in general position as described in \citet{b2}, the theoretical analysis on covariance test statistic is much more complicated and its limiting distribution has not be fully derived. \citet{b2} provides a control on the tail probability of the covariance test statistic. It will be interesting to characterize the indistinguishable region on the Lasso solution path and interpret the result of the proposed QVS method in the more general setting.  
We note that the simulation and real data analyses in the paper have design matrices that are not orthogonal. 
Compared with other popular methods, QVS shows advantages in selection accuracy and the ability to interpret its results.

\section{Appendix}
\label{Appendix}

\subsection{Proof of Theorem \ref{thm:lowerbd}.} \label{sec:proof_1}

Define $F_{m}(t) =  \frac{1}{m}\sum_{k=1}^m 1(q_k\leq t)$. 
Re-write $\hat{k}_{qvs}$ in (\ref{QVS_def}) as
\begin{eqnarray}
\hat{k}_{qvs}=m\max_{0<t<1} \{F_m(t)-t-c_m\sqrt{t(1-t)}\}.
\label{QVS_def_rew}
\end{eqnarray}
For notation convenience, define $\pi_1 = m_1/m$. Then
\begin{eqnarray}
F_{m}(t)& = & 
\frac{m_1}{m} \frac{1}{m_1} \sum_{j=1}^{m_1} 1(q_j\leq t)
+\frac{m-m_1}{m} \frac{1}{m-m_1} \sum_{j=m_1+1}^{m} 1(q_j\leq t) \nonumber\\
&\le & \pi_1 +(1-\pi_1) \frac{1}{m-m_1} \sum_{j=m_1+1}^{m} 1(q_j\leq t).
\label{Fmt_1_transf}
\end{eqnarray}
Define $U_{(1), m-s}, \ldots, U_{(m-s), m-s}$ as the increasingly ordered statistics of $m-s$ standard uniform random variables. Further, let 
\[
U_{m-m_1, m-s}(t) = \frac{1}{m-m_1} \sum_{j=m_1-s+1}^{m-s} 1(U_{(j), m-s}\leq t). 
\]
Then, $ \frac{1}{m-m_1} \sum_{j=m_1+1}^{m} 1(q_j\leq t)\overset{d}{=} U_{m-m_1, m-s}(t) $
and $F_{m}(t) \le \pi_1 +(1-\pi_1) U_{m-m_1, m-s}(t)$.
Therefore, 

\begin{eqnarray}
& & P(\hat{k}_{qvs}>m_1) \nonumber\\
& = & P(\sup_{0<t<1} \{F_m(t)-t-c_m\sqrt{t(1-t)}\}>\pi_1) \nonumber\\
&\le & P(\sup_{0<t<1} \{\pi_1+(1-\pi_1)U_{m-m_1, m-s}(t)-t-c_m\sqrt{t(1-t)}\}>\pi_1) \nonumber\\
& = & P(\sup_{0<t<1} \{(1-\pi_1)U_{m-m_1, m-s}(t)-t-c_m\sqrt{t(1-t)}\}>0) \nonumber\\
& \leq & 
P(\sup_{0<t<1} \{(1-\pi_1)U_{m-m_1, m-s}(t)-U_m(t)\}>0) +
P(\sup_{0<t<1} \{U_m(t)-t-c_m\sqrt{t(1-t)}\}>0)
\label{G_mid}
\end{eqnarray}

For the first term in (\ref{G_mid}), we have 
\begin{eqnarray*}
	(1-\pi_1)U_{m-m_1, m-s}(t)-U_m(t) & = & \frac{1}{m}\left (
	\sum_{j=m_1-s+1}^{m-s}  1(U_{(j),m-s}\leq t) - \sum_{j=1}^m 1(U_j\leq t)
	\right ) \nonumber \\
	& \le &\frac{1}{m}\left (
	\sum_{j=1}^{m-s}  1(U_{(j),m-s}\leq t) - \sum_{j=1}^m 1(U_j\leq t)
	\right )\nonumber \\
	& \overset{d}{=} & \frac{1}{m}\left (
	\sum_{j=1}^{m-s}  1(U'_j\leq t) - \sum_{j=1}^m 1(U_j\leq t)
	\right ),
\end{eqnarray*}
where  $\{U'_j\}_{j=1}^{m-s}$ is a sequences of independent standard uniform random variables, which are independent of $\{U_j\}_{j=1}^m$. By Chebyshev's inequality and $s=o(m)$, 
\begin{equation*} 
P\left(\frac{1}{m}\left (
\sum_{j=1}^{m-s}  1(U'_j\leq t) - \sum_{j=1}^m 1(U_j\leq t)
\right ) > 0 \right) = o(1)
\end{equation*}
for any $t>0$. Then 
\begin{equation} \label{1}
P(\sup_{0<t<1} \{(1-\pi_1)U_{m-m_1, m-s}(t)-U_m(t)\}>0) = o(1). 
\end{equation}

Now, consider the second term in (\ref{G_mid}). By the definition of the bounding sequence $c_m$,
\begin{eqnarray*}
	P\left(\max_{0<t<1} \frac{U_m(t)-t}{\sqrt{t(1-t)}} >c_m\right)=\alpha_m.\label{alphaden}
\end{eqnarray*}
Further,
\[
P\left(\max_{0<t<1} \frac{U_m(t)-t-c_m\sqrt{t(1-t)}}{\sqrt{t(1-t)}} >0\right) \le P\left(\max_{0<t<1} \frac{U_m(t)-t}{\sqrt{t(1-t)}} >c_m\right) =\alpha_m.
\]
And for every $t\in (0,1)$,
\[
\frac{U_m(t)-t-c_m\sqrt{t(1-t)}}{\sqrt{t(1-t)}}>U_m(t)-t-c_m\sqrt{t(1-t)}.
\]
The above implies that
\begin{equation} \label{2}
P\left(\max_{0<t<1} \{U_m(t)-t-c_m\sqrt{t(1-t)}\} >0\right)\leq \alpha_m = o(1).
\end{equation}

Combining (\ref{G_mid}) with (\ref{1}) and (\ref{2}) gives 
$P(\hat{k}_{qvs}>m_1) \to 0$ as $m\to\infty$. \qed

\subsection{Proof of Theorem \ref{thm:upperbd_m0}.} \label{sec:proof_2}

By the construction of $\hat{k}_{qvs}$ in (\ref{QVS_def}),  
\begin{eqnarray} \label{5.1}
\frac{\hat{k}_{qvs}}{m_0} - 1 & = & \max_{1\leq k\leq m/2} \left\{
\frac{k}{m_0}- 1 - \frac{m}{m_0} q_k- \frac{m}{m_0}c_m\sqrt{q_k(1-q_k)}\right\} \nonumber \\
& \ge & \frac{m_0 + 1}{m_0}- 1 - \frac{m}{m_0} q_{m_0+1}- \frac{m}{m_0}c_m\sqrt{q_{m_0+1}} \nonumber \\
& > & - \frac{m}{m_0} q_{m_0+1} - \frac{m}{m_0}c_m\sqrt{q_{m_0+1}},
\end{eqnarray}
where the second step above is by taking $k=m_0+1$. 

Recall the positions of the noise variables $a_1 , \ldots, a_{m-s}$ and $a_1 = m_0+1$.  Under condition (\ref{cond:T}), R{\'e}nyi representation gives $q_{m_0+1} = q_{a_1}  \le  U_{(1), m-s}$, where  $U_{(1), m-s} \sim Beta(1, m-s)$. Therefore, given $s = o(m)$ and $m_0 \gg \sqrt{\log m}$ with probability tending to 1,
\begin{equation} \label{5.2}
\frac{m}{m_0} q_{m_0+1} = \frac{m}{m_0} q_{a_1} = O_p(\frac{1}{m_0}) = o_p(1). 
\end{equation}
On the other hand, it has been shown in \cite{e1} that $c_m = O(\sqrt{\log m}/ \sqrt{m})$.  Then, by conditions $s = o(m)$ and $m_0 \gg \sqrt{\log m}$ with probability tending to 1, we have 
\begin{equation} \label{5.3}
\frac{m}{m_0}c_m\sqrt{q_{m_0+1}} = \frac{m}{m_0}c_m\sqrt{q_{a_1}} = O_p(\frac{\sqrt{\log m}}{m_0}) = o_p(1). 
\end{equation}
Combining (\ref{5.1}) - (\ref{5.3}) gives (\ref{5}). \qed

\subsection{Proof of Theorem \ref{thm:upperbd_m1}.} \label{sec:proof_3}

Given the lower bound result in Theorem \ref{thm:lowerbd}, it is enough to show
\begin{equation} \label{3}
P\{\hat{k}_{qvs} > (1-\epsilon)m_1\} \to 1
\end{equation}
for an arbitrarily small constant $\epsilon>0$ as $m\to\infty$.

By the construction of $\hat{k}_{qvs}$ in (\ref{QVS_def}),  
\begin{eqnarray} \label{3.1}
\frac{\hat{k}_{qvs}}{m_1} - 1 & = & \max_{1\leq k\leq m/2} \left\{
\frac{k}{m_1}- 1 - \frac{m}{m_1} q_k- \frac{m}{m_1}c_m\sqrt{q_k(1-q_k)}\right\} \nonumber \\
& \ge & \frac{m_1 + 1}{m_1}- 1 - \frac{m}{m_1} q_{m_1+1}- \frac{m}{m_1}c_m\sqrt{q_{m_1+1}} \nonumber \\
& > & - \frac{m}{m_1} q_{m_1+1} - \frac{m}{m_1}c_m\sqrt{q_{m_1+1}},
\end{eqnarray}
where the second step above is by taking $k=m_1+1$. 
Recall the positions of the noise variables $a_1 , \ldots, a_{m-s}$ and $a_{m_1-s+1} = m_1+1$.  By condition (\ref{cond:T}) and R{\'e}nyi representation, $q_{m_1+1} = q_{a_{m_1-s+1}}  \overset{d}{=} U_{(m_1-s+1), m-s}$, where  $U_{(m_1-s+1), m-s} \sim Beta(m_1-s+1, m-s)$. 
Therefore, given $s = o(m)$ and $(m_1-s)/m_1 = o_p(1)$,
\begin{equation} \label{3.2}
\frac{m}{m_1} q_{m_1+1} = \frac{m}{m_1} q_{a_{m_1-s+1}} = O_p(\frac{m_1-s}{m_1}) = o_p(1). 
\end{equation}
On the other hand, under conditions $s = o(m)$ and $m_1 \gg \sqrt{\log m}$ and $(m_1-s)/ m_1 \ll 1$ with probability tending to 1,  similar arguments as those leading to (\ref{5.3}) gives 
\begin{equation} \label{3.3}
\frac{m}{m_1}c_m\sqrt{q_{m_1+1}} = \frac{m}{m_1}c_m\sqrt{q_{a_{m_1-s+1}}} = O_p(\frac{\sqrt{(m_1-s)\log m }}{m_1}) = o_p(1). 
\end{equation}
Combining (\ref{3.1}) - (\ref{3.3}) gives (\ref{3}). \qed

\subsection{Additional simulation results} \label{sec:add_sim}

\begin{figure}[htpb]
	\includegraphics[width=0.32\textwidth,height=0.32\textheight]{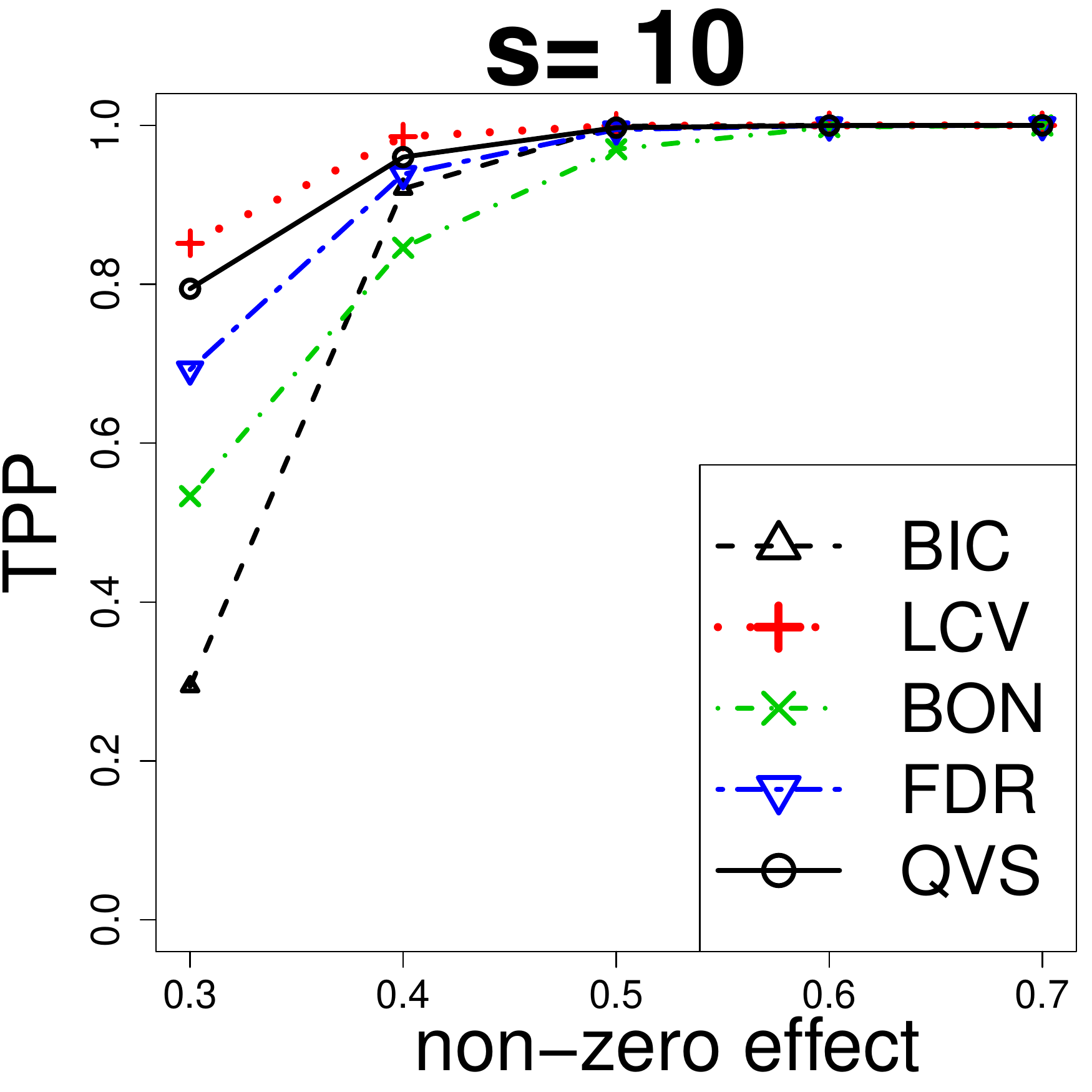}
	\includegraphics[width=0.32\textwidth,height=0.32\textheight]{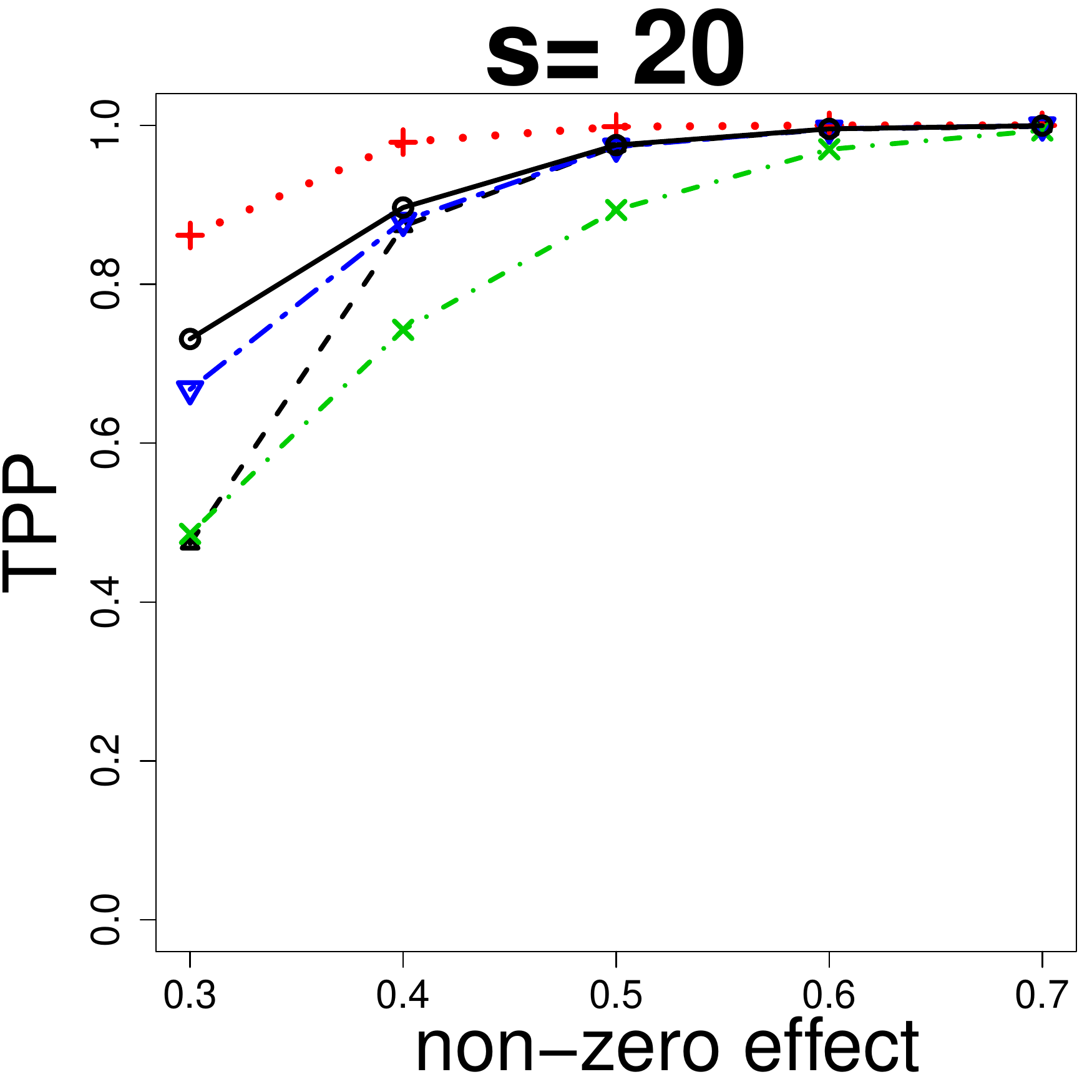}
	\includegraphics[width=0.32\textwidth,height=0.32\textheight]{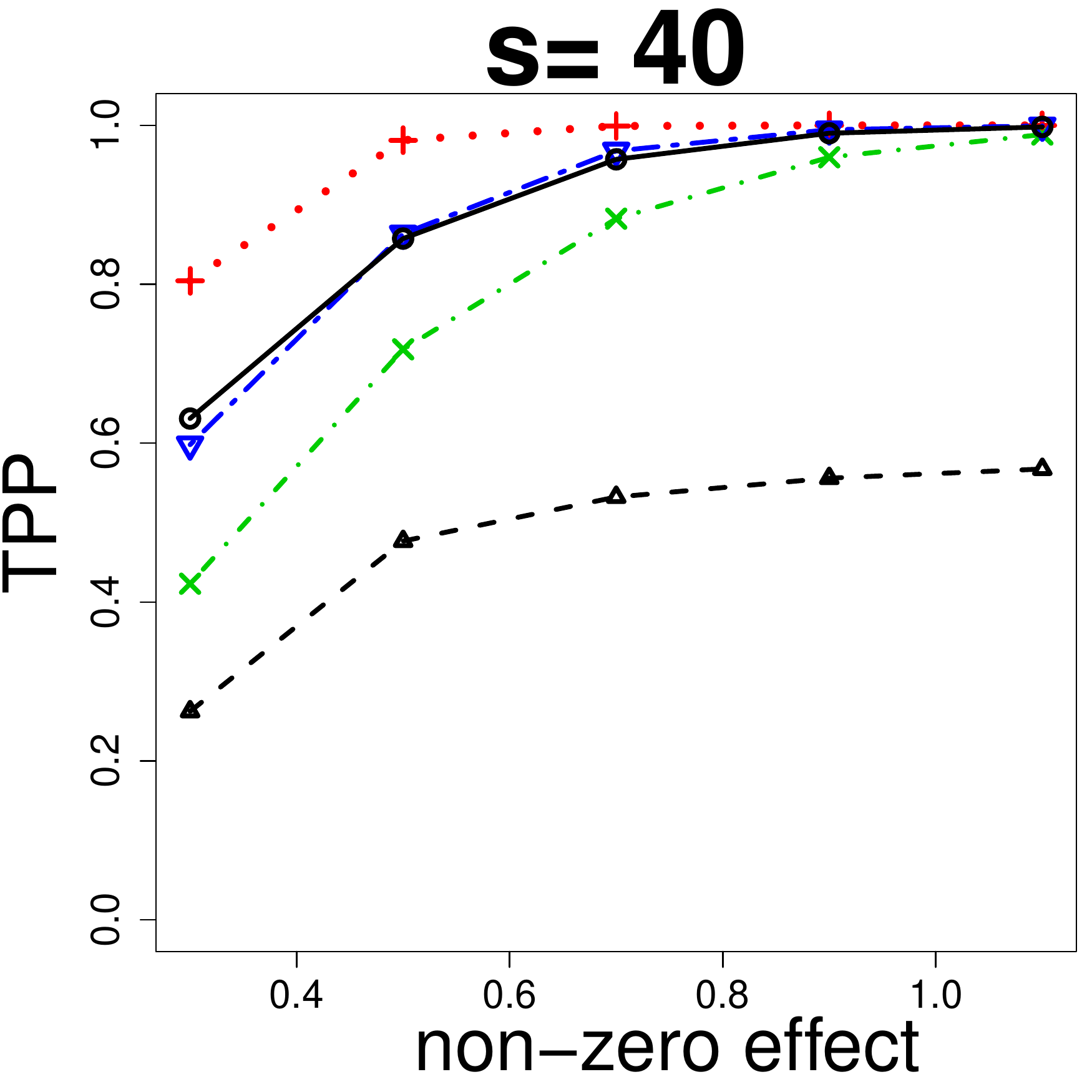}
	\includegraphics[width=0.32\textwidth,height=0.32\textheight]{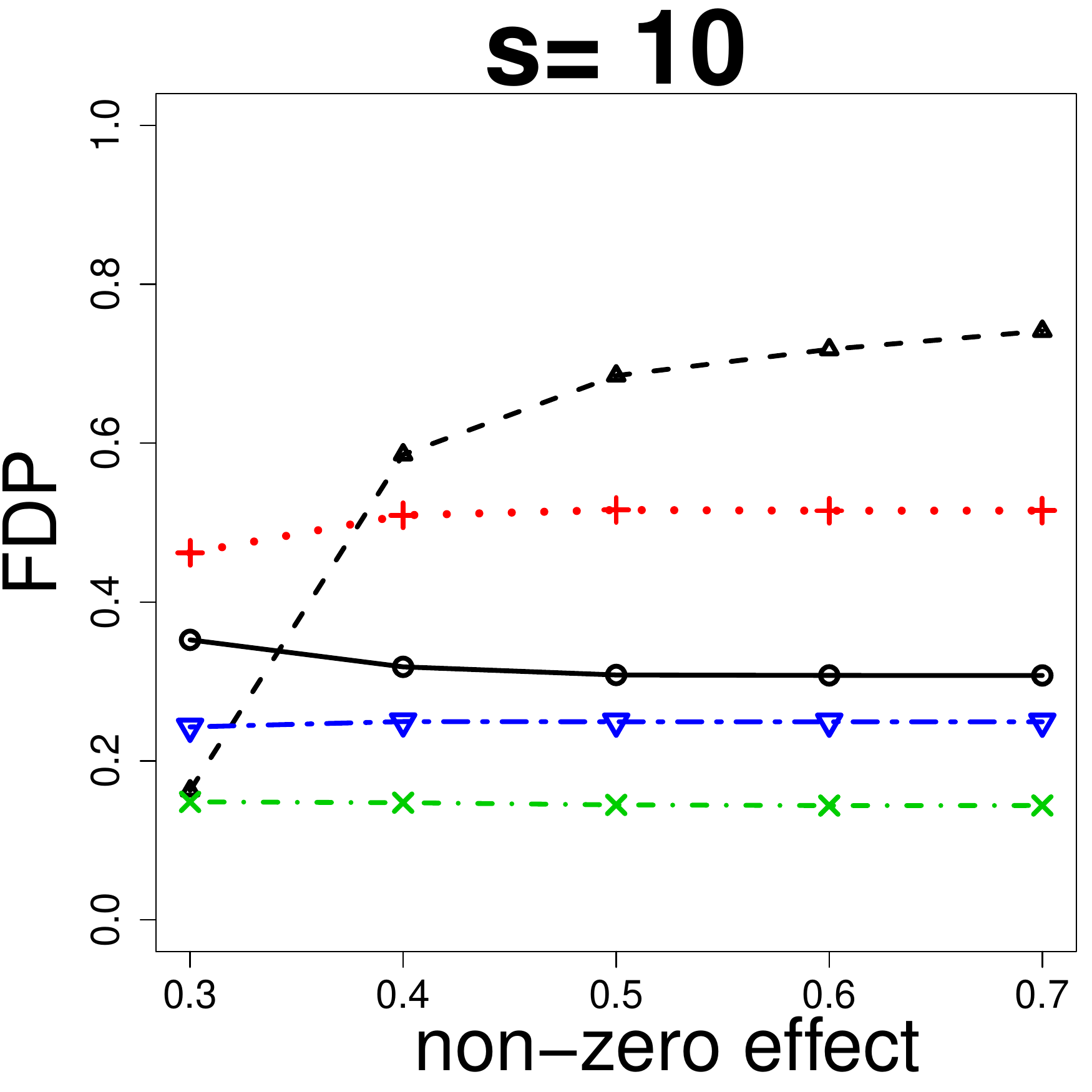}
	\includegraphics[width=0.32\textwidth,height=0.32\textheight]{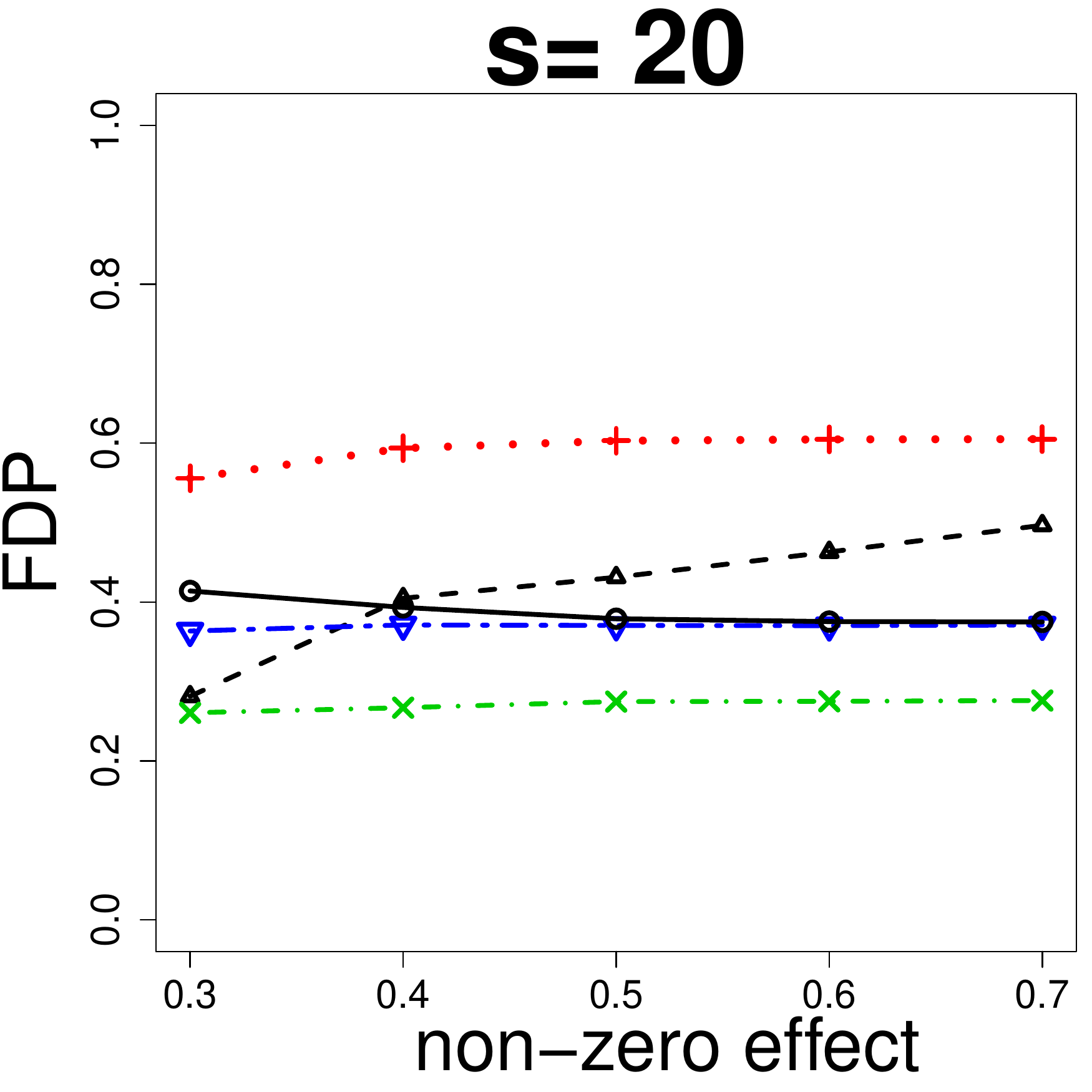}
	\includegraphics[width=0.32\textwidth,height=0.32\textheight]{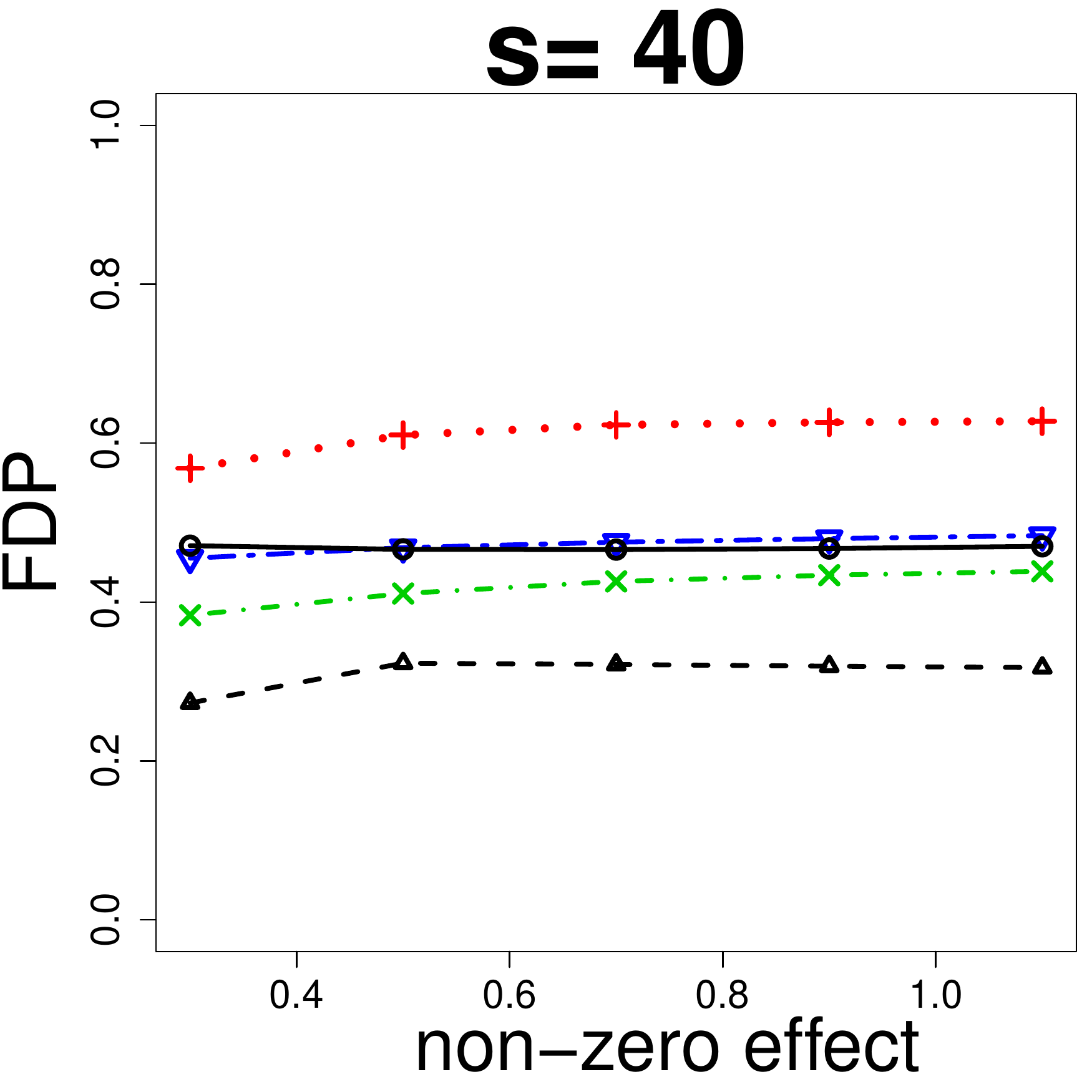}
	\includegraphics[width=0.32\textwidth,height=0.32\textheight]{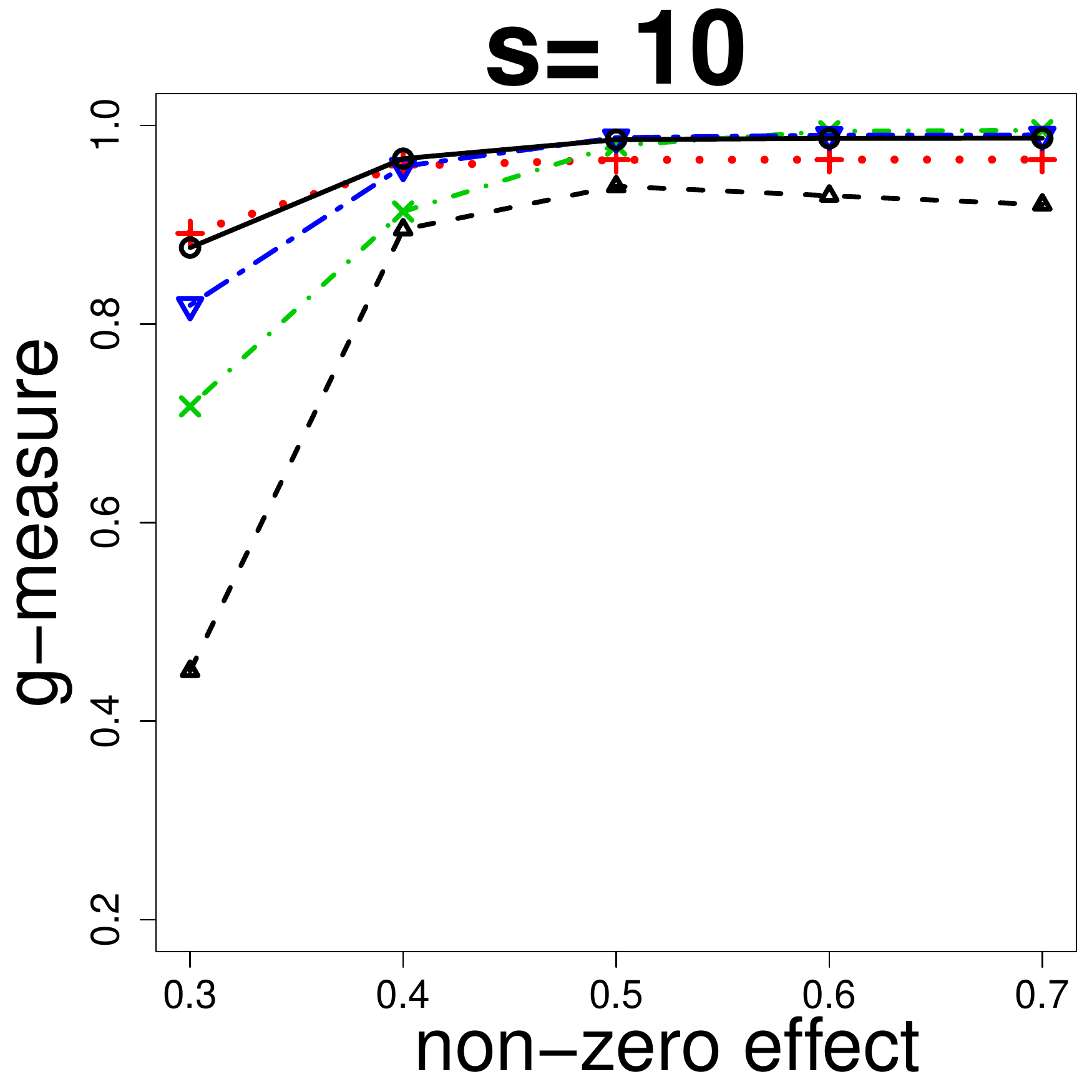}
	\includegraphics[width=0.32\textwidth,height=0.32\textheight]{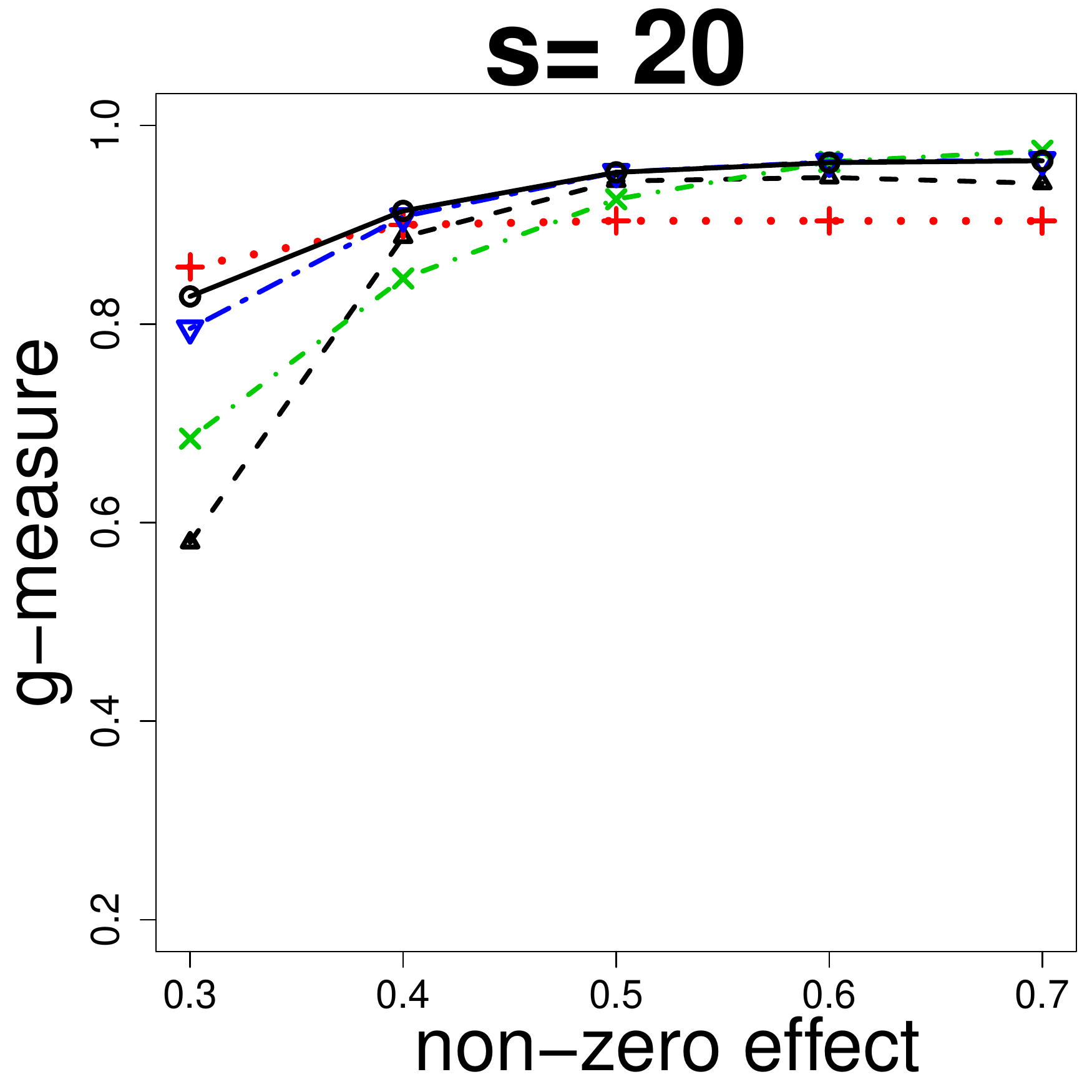}
	\includegraphics[width=0.32\textwidth,height=0.32\textheight]{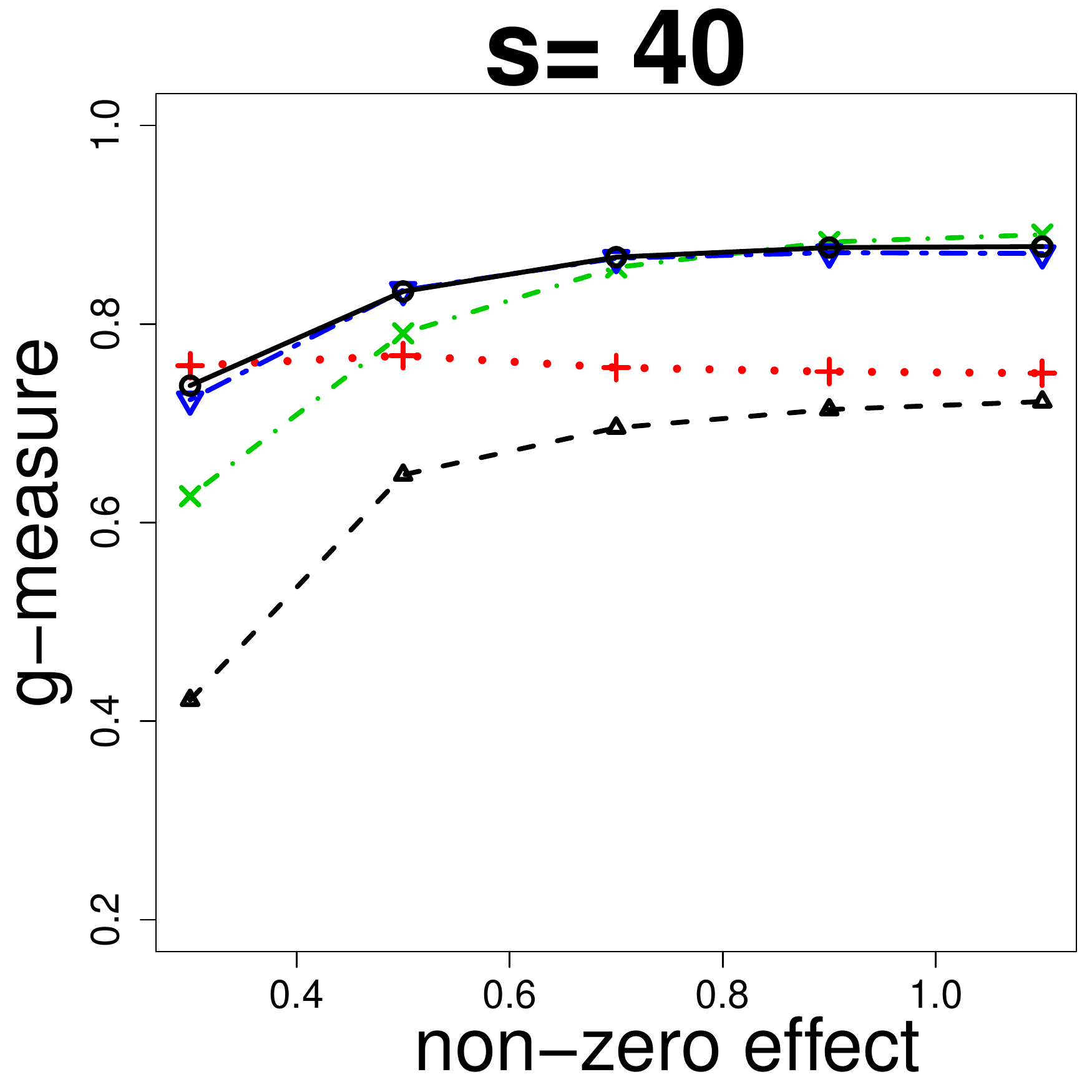}
	\caption{Comparisons of QVS with other methods when $p=400$, $\ms{\Sigma}=(0.5^{|i-j|})_{i=1,\cdots,p;~j=1,\cdots,p}$, and $n=200$.}
	\label{p400}.
\end{figure}

\begin{figure}[htpb]
	\includegraphics[width=0.32\textwidth,height=0.32\textheight]{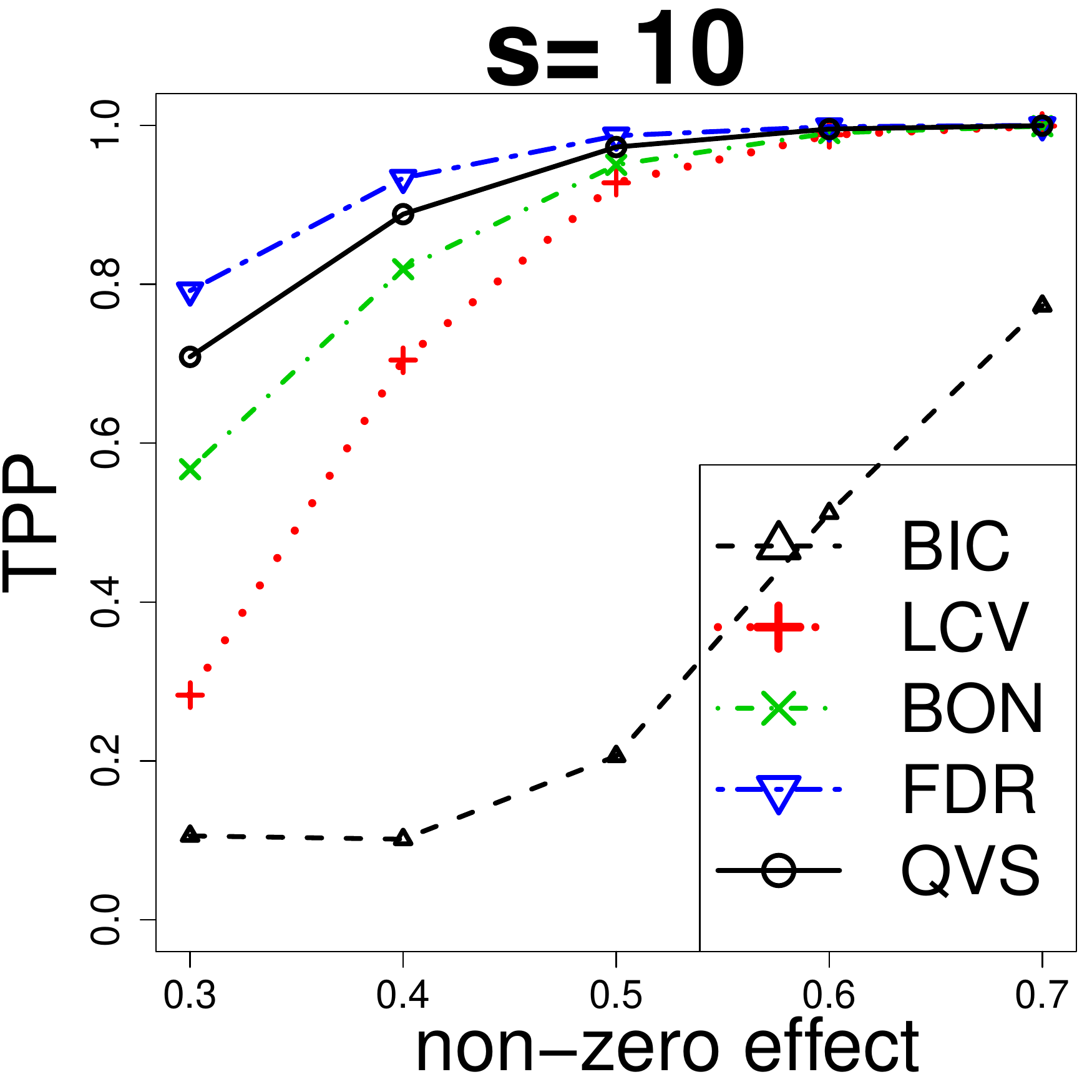}
	\includegraphics[width=0.32\textwidth,height=0.32\textheight]{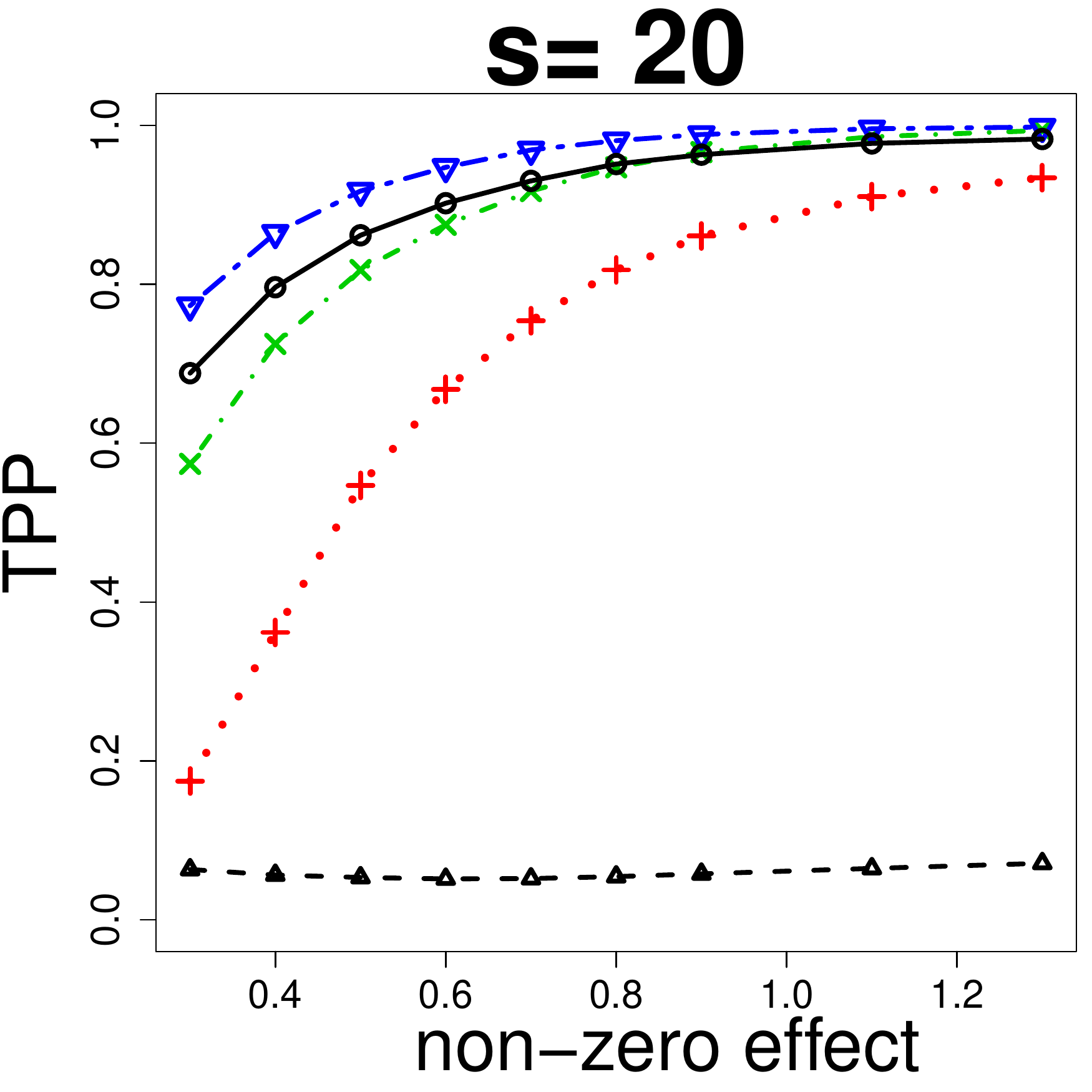}
	\includegraphics[width=0.32\textwidth,height=0.32\textheight]{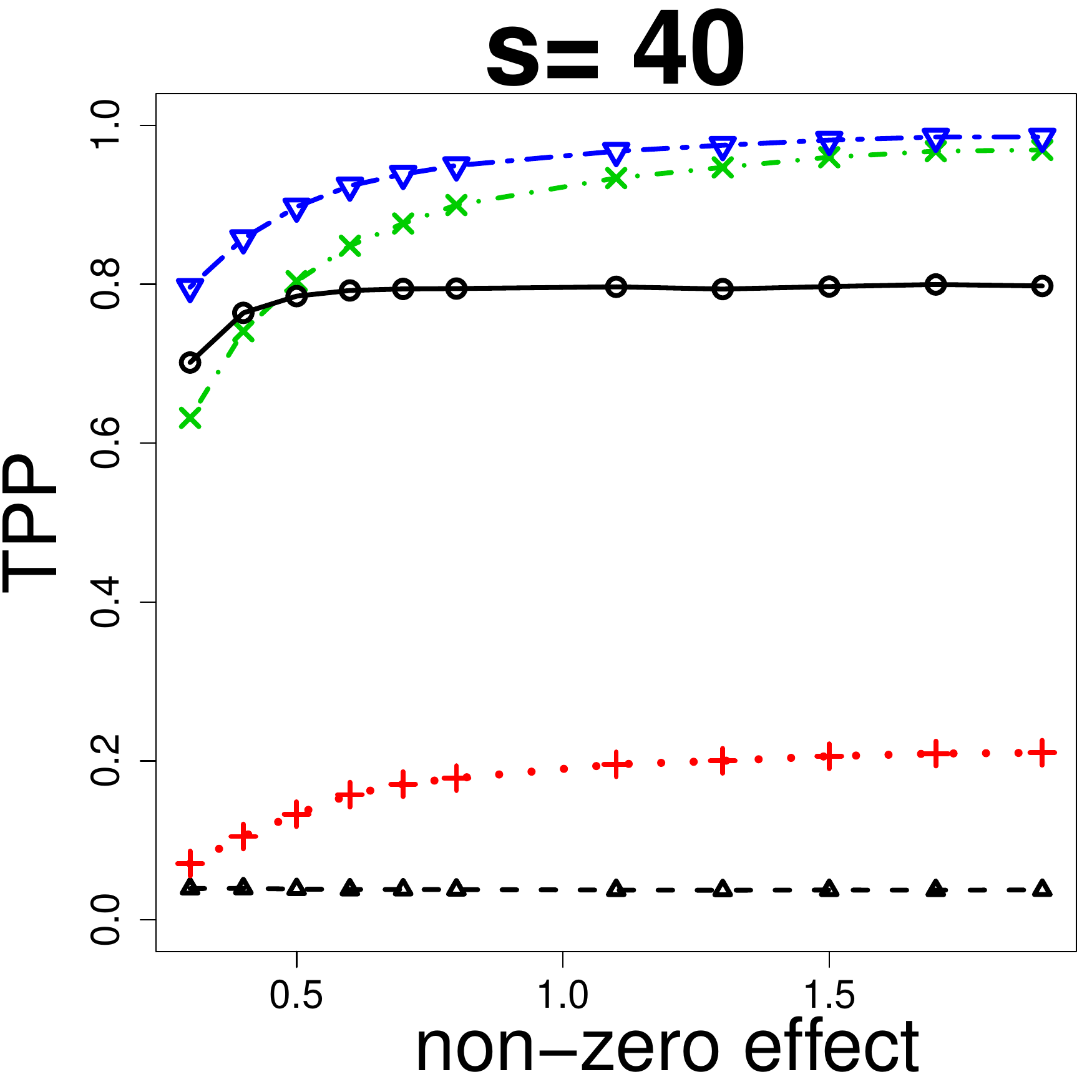}
	\includegraphics[width=0.32\textwidth,height=0.32\textheight]{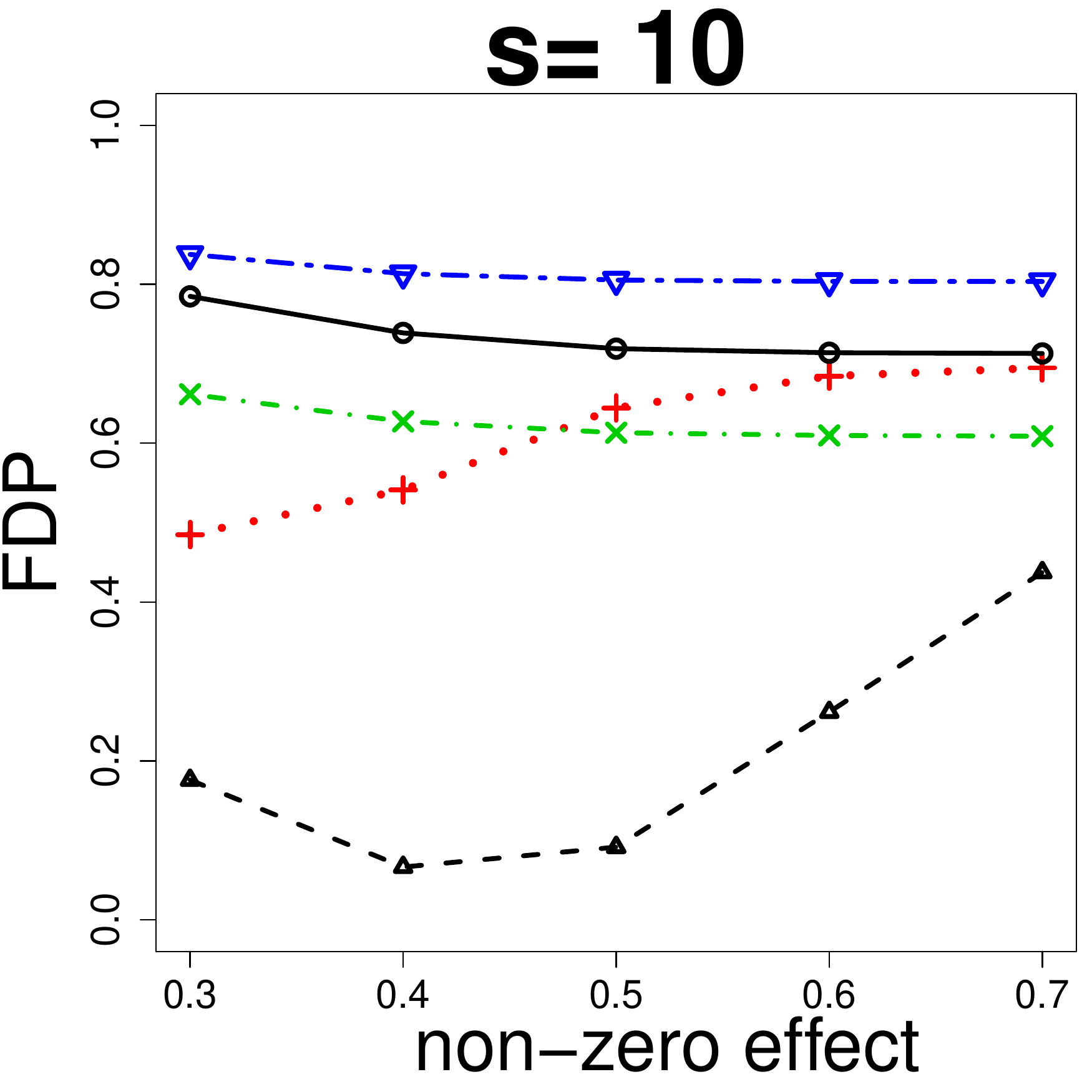}
	\includegraphics[width=0.32\textwidth,height=0.32\textheight]{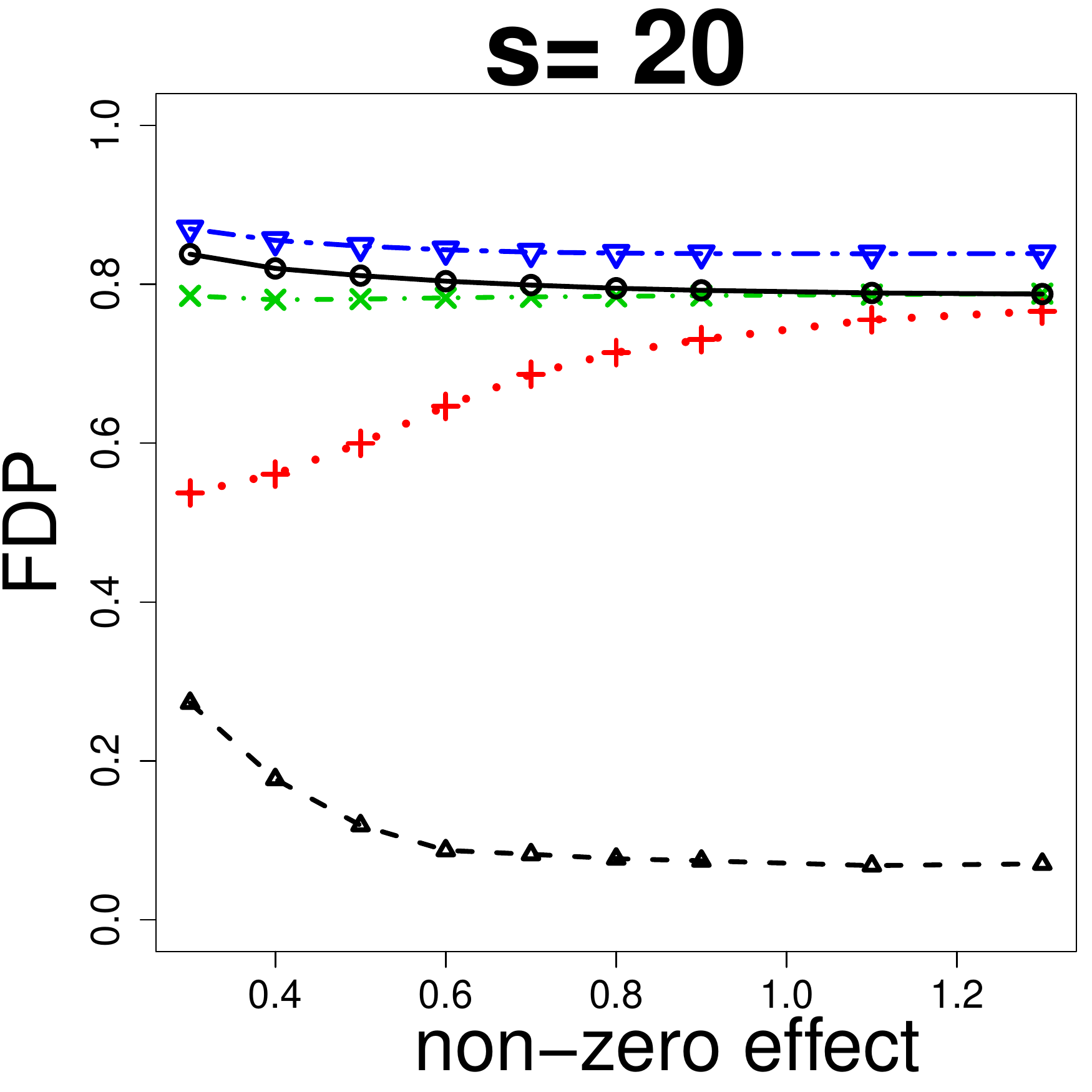}
	\includegraphics[width=0.32\textwidth,height=0.32\textheight]{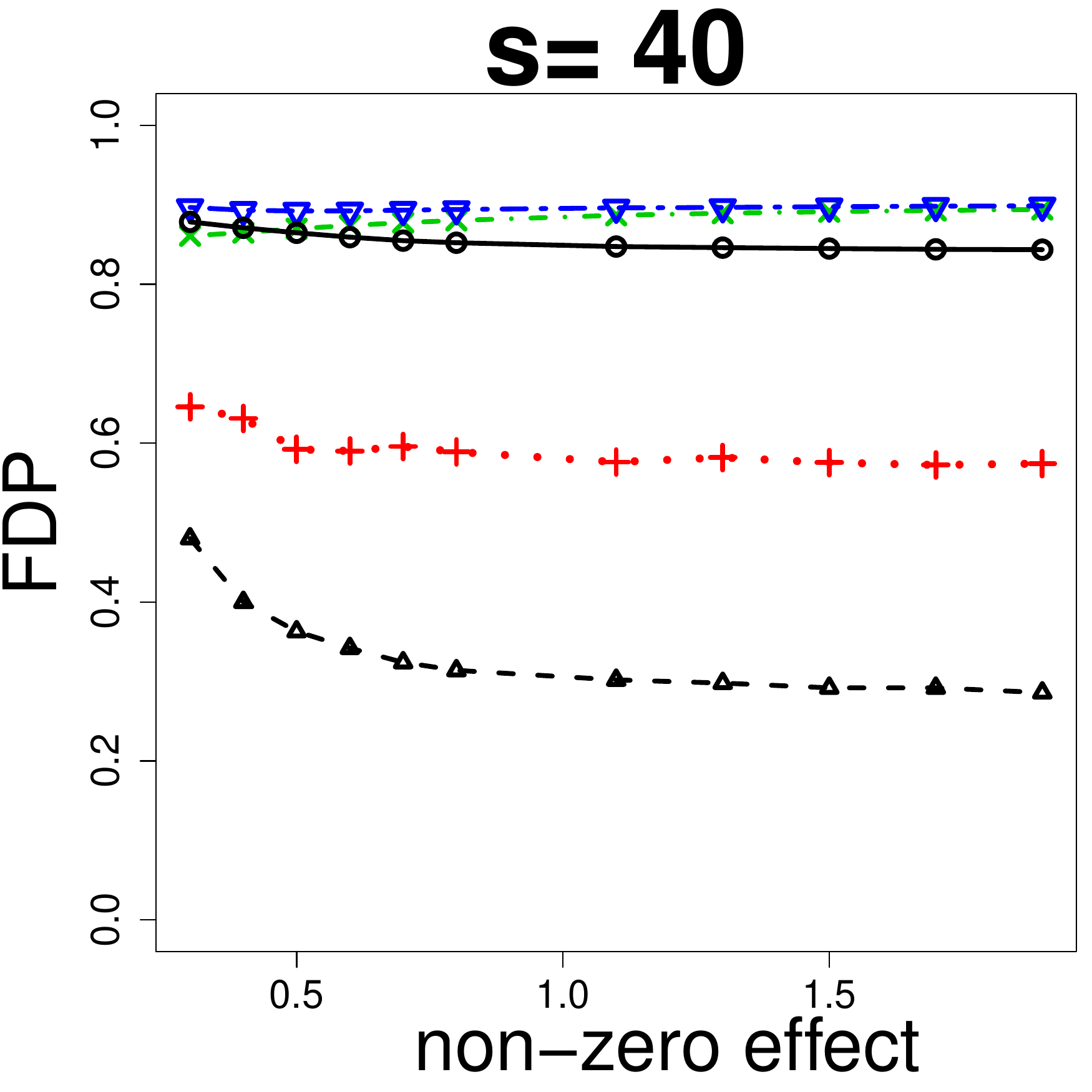}
	\includegraphics[width=0.32\textwidth,height=0.32\textheight]{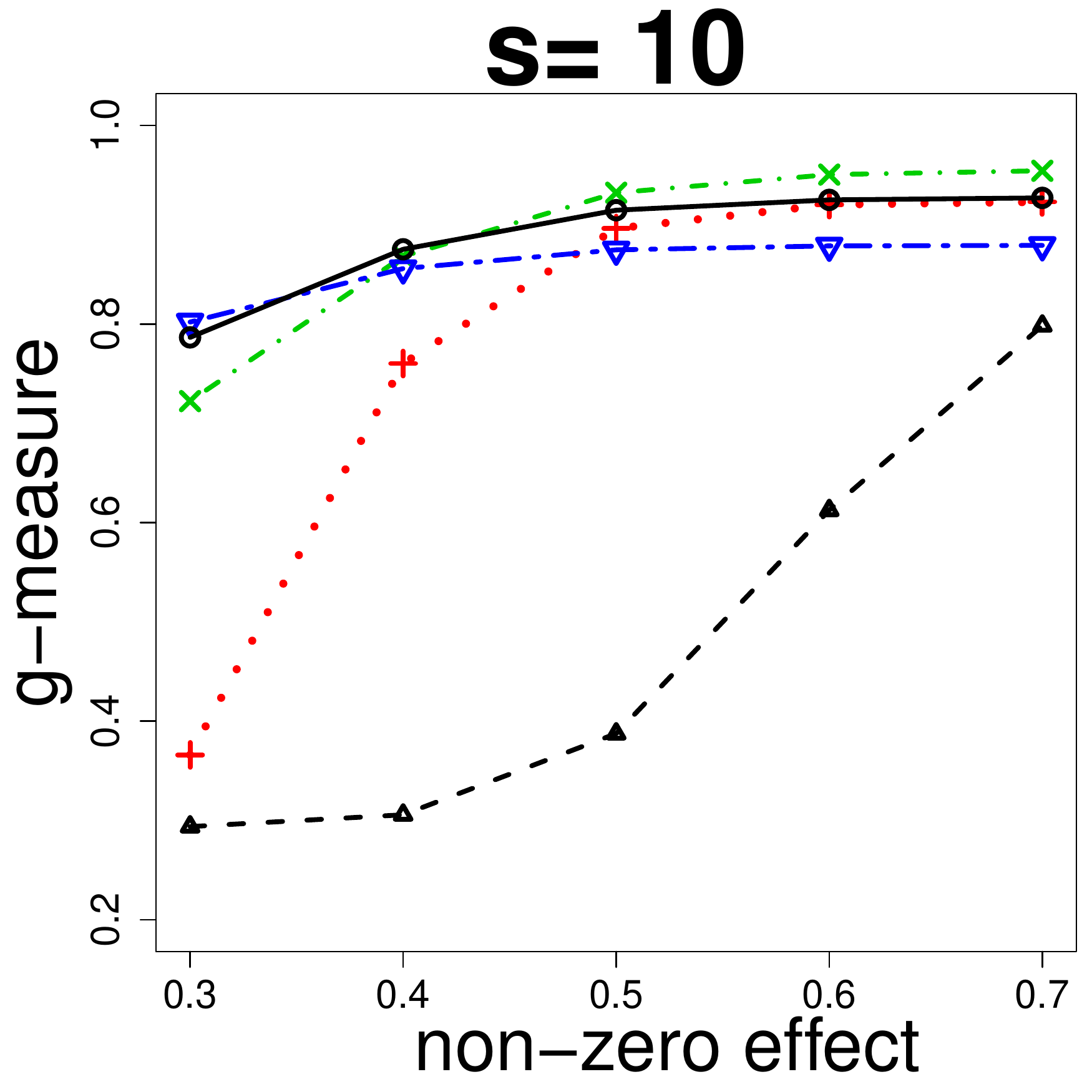}
	\includegraphics[width=0.32\textwidth,height=0.32\textheight]{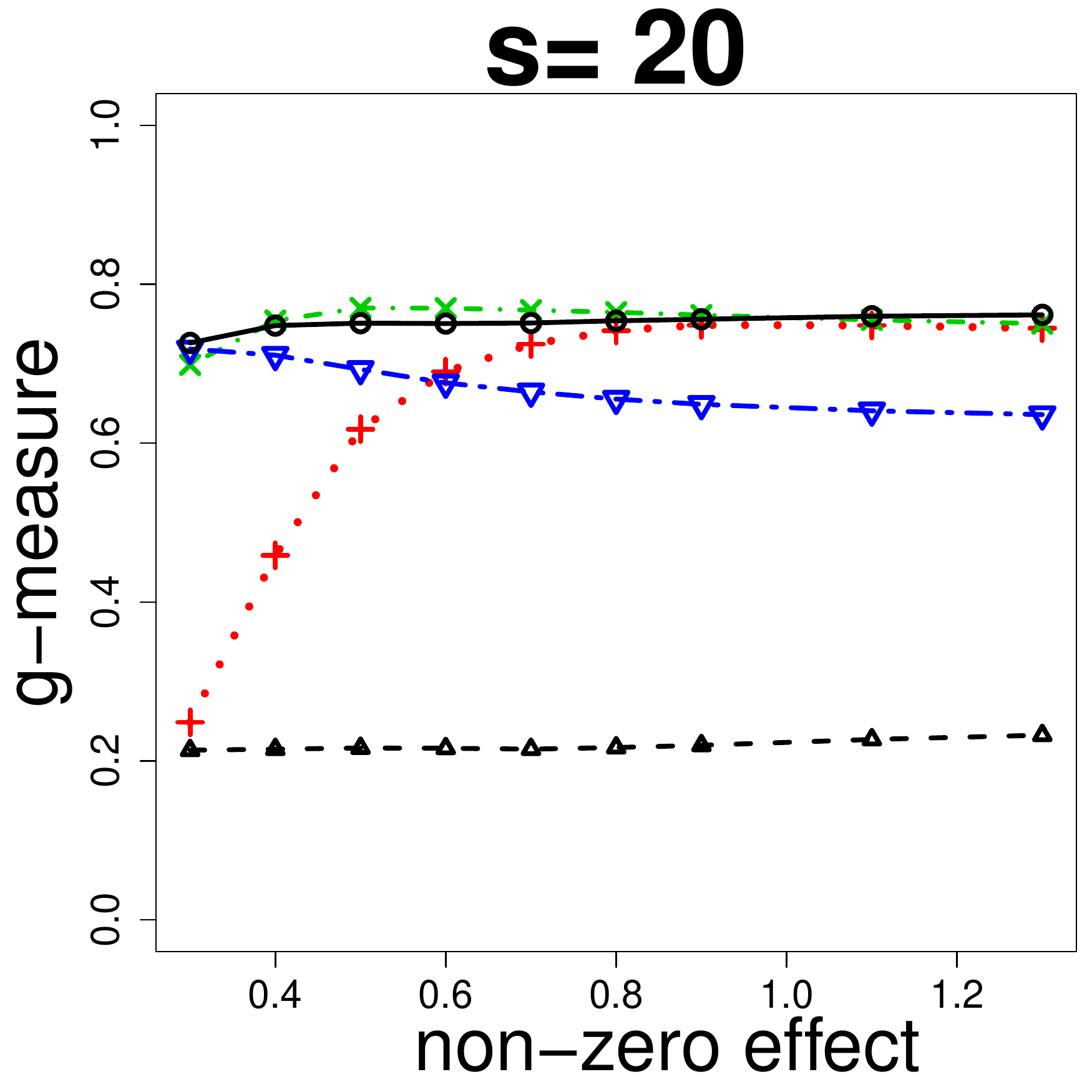}
	\includegraphics[width=0.32\textwidth,height=0.32\textheight]{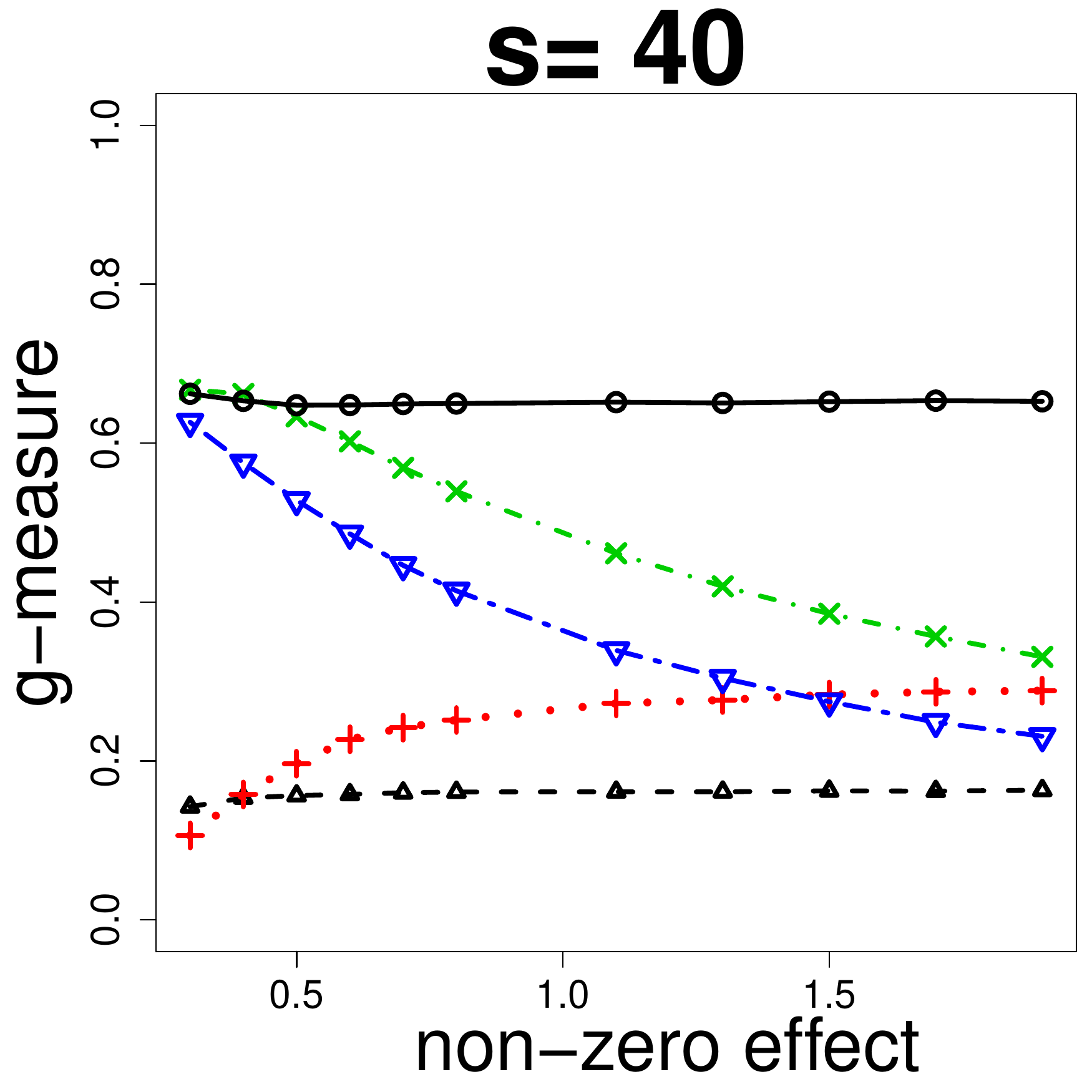}
	\caption{Comparisons of QVS with other methods when $p=10000$, $\ms{\Sigma}=(0.5^{|i-j|})_{i=1,\cdots,p;~j=1,\cdots,p}$, and $n=200$. }
	\label{p10000}
\end{figure}

We consider more model settings with $n=200$ and $Cov(\mb{X}) = \ms{\Sigma}=(0.5^{|i-j|})_{i=1,\cdots,p;~j=1,\cdots,p}$. The dimension varies with $p=400$ and $10000$. The number of non-zero coefficients and the non-zero effect also vary from case to case.
Figure \ref{p400} and \ref{p10000} summarize the mean values of TPP, FDP, and g-measure for different methods. It can be seen that in these settings, Q-BON does not control family-wise error under the nominal level of 0.05, and Q-FDR does not control FDR under the nominal level of 0.05. QVS generally outperforms other methods in g-measure.

\bibliographystyle{wb_stat}
\bibliography{Reference_path}

\begin{thebibliography}{16}
\newcommand{\enquote}[1]{`#1'}
\providecommand{\natexlab}[1]{#1}
\expandafter\ifx\csname urlstyle\endcsname\relax
  \providecommand{\doi}[1]{doi:\discretionary{}{}{}#1}\else
  \providecommand{\doi}{doi:\discretionary{}{}{}\begingroup
  \urlstyle{rm}\Url}\fi

\bibitem[{Barber \& Cand\`{e}s(2015)}]{Barber15}
Barber, RF \& Cand\`{e}s, EJ (2015), \enquote{Controlling the false discovery
  rate via knockoffs,} \emph{Ann. Statist.}, \textbf{43}(5), pp. 2055--2085.

\bibitem[{Benjamini \& Hochberg(1995)}]{wsc19}
Benjamini, Y \& Hochberg, Y (1995), \enquote{Controlling the false discovery
  rate: A practical and powerful approach to multiple testing,} \emph{Journal
  of the Royal Statistical Society: Series B}, \textbf{57}(1), pp. 289--300.

\bibitem[{Bogdan et~al.(2015)Bogdan, van~den Berg, Sabatti, Su \&
  Cand{\'e}s}]{Bogdan15}
Bogdan, M, van~den Berg, E, Sabatti, C, Su, W \& Cand{\'e}s, E (2015),
  \enquote{{SLOPE} --- adaptive variable selection via convex optimization,}
  \emph{Ann. Appl. Stat.}, \textbf{9}(3), pp. 1103--1140.

\bibitem[{Bradic et~al.(2011)Bradic, Fan \& Wang}]{g1}
Bradic, J, Fan, J \& Wang, W (2011), \enquote{Penalized composite
  quasi-likelihood for unltahigh-dimensional variable selection,} \emph{Journal
  of the royal statistical society: series B (statistical methodology)},
  \textbf{73}(3), pp. 325--349.

\bibitem[{Efron et~al.(2004)Efron, Hastie, Johnstone \& Tibshirani}]{b5}
Efron, B, Hastie, T, Johnstone, I \& Tibshirani, R (2004), \enquote{Least angle
  regression,} \emph{The Annals of Statistics}, \textbf{32}(2), pp. 407--499.

\bibitem[{G'sell et~al.(2016)G'sell, Wager, Chouldechova \&
  Tibshirani}]{Gsell16}
G'sell, M, Wager, S, Chouldechova, A \& Tibshirani, R (2016),
  \enquote{Sequential selection procedures and false discovery rate control,}
  \emph{J. R. Statist. Soc. Ser. B}, \textbf{78}(2), pp. 423--444.

\bibitem[{Lee et~al.(2016)Lee, Sun, Sun \& Taylor}]{Lee16}
Lee, J, Sun, D, Sun, Y \& Taylor, J (2016), \enquote{Exact post-selection
  inference, with application to the lasso,} \emph{Ann. Statist.},
  \textbf{44}(3), pp. 907--927.

\bibitem[{Lockhart et~al.(2014)Lockhart, Taylor, Tibshirani \& Tibshirani}]{b2}
Lockhart, R, Taylor, J, Tibshirani, R \& Tibshirani, R (2014), \enquote{A
  significance test for the lasso,} \emph{The Annals of Statistics},
  \textbf{42}(2), pp. 413--468.

\bibitem[{Meinshausen \& Buhlmann(2005)}]{h1}
Meinshausen, N \& Buhlmann, P (2005), \enquote{Lower bounds for the number of
  false null hypotheses for multiple testing of associations under general
  dependence structures,} \emph{Biometrika}, \textbf{92}(4), pp. 893--907.

\bibitem[{Meinshausen \& Rice(2006)}]{e1}
Meinshausen, N \& Rice, J (2006), \enquote{Estimating the proportion of false
  null hypotheses among a large number of independently tested hypotheses,}
  \emph{The Annals of Statistics}, \textbf{34}(1), pp. 373 -- 393.

\bibitem[{Powers(2011)}]{Powers11}
Powers, D (2011), \enquote{Evaluation: from precision, recall and f-measure to
  roc, informedness, markedness {\&} correlation.} \emph{J Machine Learning
  Technologies}, \textbf{2}, pp. 37--63.

\bibitem[{Su et~al.(2017)Su, M. \& Candes}]{Su17}
Su, W, M., B \& Candes, E (2017), \enquote{False discoveries occur early on the
  lasso path,} \emph{Ann. Statist.}, \textbf{To appear}.

\bibitem[{Tibshirani(1996)}]{ff3}
Tibshirani, R (1996), \enquote{Regression shrinkage and selection via the
  lasso,} \emph{Journal of the Royal Statistical Society: Series B},
  \textbf{58}, pp. 267--288.

\bibitem[{Van De~Geer et~al.(2014)Van De~Geer, Buhlmann, Ritov \& Dezeure}]{b3}
Van De~Geer, S, Buhlmann, P, Ritov, Y \& Dezeure, R (2014), \enquote{On
  asymptotically optimal confidence regions and tests for high-dimensional
  models,} \emph{The Annals of Statistics}, \textbf{42}(3), pp. 1166--1202.

\bibitem[{Wainwright(2009)}]{Wainwright:2009}
Wainwright, M (2009), \enquote{Sharp thresholds for high-dimensional and noisy
  sparsity recovery using ${\ell}_1$-constrained quadratic programming
  (lasso),} \emph{IEEE Transactions on Information Theory}, \textbf{55}(5), pp.
  2183--2202.

\bibitem[{Zhang \& Zhang(2014)}]{b20}
Zhang, C \& Zhang, SS (2014), \enquote{Confidence intervals for low dimensional
  parameters in high dimensional linear models,} \emph{Journal of the Royal
  Statistical Society: Series B}, \textbf{76}(1), pp. 217--242.

\end{thebibliography}

\end{document}